\documentclass[twocolumn,showpacs,preprintnumbers,nofootinbib,prd,
superscriptaddress,10pt]{revtex4-1}

\usepackage{lmodern}
\usepackage{amsmath,amssymb}
\usepackage[normalem]{ulem}
\usepackage{textcomp}
\usepackage{hyperref}
\usepackage{bm}
\usepackage{graphicx}
\usepackage{psfrag}
\usepackage[usenames,dvipsnames]{xcolor}
\usepackage[utf8]{inputenc}

\definecolor{darkgreen}{rgb}{0,0.5,0}

\hypersetup{
    bookmarks=true,         % show bookmarks bar?
    unicode=false,          % non-Latin characters in Acrobat???s bookmarks
    pdftoolbar=true,        % show Acrobat???s toolbar?
    pdfmenubar=true,        % show Acrobat???s menu?
    pdffitwindow=false,     % window fit to page when opened
    pdfstartview={FitH},    % fits the width of the page to the window
	pdftitle={Gravitational-wave amplitudes for compact binaries in eccentric 
		orbits at the third post-Newtonian order: Tail contributions and 
		post-adiabatic corrections},
    pdfauthor={Y.~Boetzel, C.~K.~Mishra, G.~Faye, A.~Gopakumar, B.~Iyer},
    pdfsubject={Gravitational Waves},   % subject of the document
    pdfkeywords={gravitational waves, eccentric compact binaries, general 
		relativity},
    pdfnewwindow=true,      % links in new window
    colorlinks=true,       % false: boxed links; true: colored links
    linkcolor=red,          % color of internal links
    citecolor=cyan,        % color of links to bibliography
    filecolor=magenta,      % color of file links
    urlcolor=darkgreen,           % color of external links
    linktocpage=true
}

\allowdisplaybreaks[4]

\newcommand{\bigO}{\mathcal{O}}
\newcommand{\la}{\langle}
\newcommand{\ra}{\rangle}
\newcommand{\sign}[1]{\text{sign}(#1)}

\newcommand{\pd}{\partial}

\newcommand{\eg}{\gamma_\textnormal{E}}
\newcommand{\inst}{\textnormal{inst}}
\newcommand{\hered}{\textnormal{hered}}
\newcommand{\tail}{\textnormal{tail}}
\newcommand{\mem}{\textnormal{mem}}
\newcommand{\memdc}{\textnormal{DC}}
\newcommand{\memosc}{\textnormal{osc}}
\newcommand{\post}{\textnormal{post-ad}}
\newcommand{\ud}{\mathrm{d}}
\newcommand{\ui}{\mathrm{i}}
\newcommand{\ue}{\mathrm{e}}
\newcommand{\TT}{\textnormal{TT}}
\newcommand{\PTT}{\mathcal{P}}

\newcommand{\xb}{\bar{x}}
\newcommand{\xp}{\tilde{x}}
\newcommand{\eb}{\bar{e}}
\newcommand{\ep}{\tilde{e}}
\newcommand{\clb}{\bar{c}_l}
\newcommand{\clp}{\tilde{c}_l}
\newcommand{\clamb}{\bar{c}_\lambda}
\newcommand{\clamp}{\tilde{c}_\lambda}
\newcommand{\lb}{\bar{l}}
\newcommand{\lp}{\tilde{l}}
\newcommand{\lab}{\bar{\lambda}}
\newcommand{\lap}{\tilde{\lambda}}
\newcommand{\ub}{\bar{u}}
\newcommand{\up}{\tilde{u}}
\newcommand{\vb}{\bar{v}}
\newcommand{\vp}{\tilde{v}}

\graphicspath{{./figures/}}

\begin{document}

\title{Gravitational-wave amplitudes for compact binaries in eccentric orbits 
at the third post-Newtonian order: Tail contributions and post-adiabatic 
corrections}

\date{\today}

\author{Yannick Boetzel}
\affiliation{Physik-Institut, Universit\"at Z\"urich, 8057 Z\"urich, 
Switzerland}

\author{Chandra Kant Mishra} \affiliation{Department of Physics, Indian 
	Institute of Technology, Madras, Chennai 600036, India}

\author{Guillaume Faye}
\affiliation{$\mathcal{G}\mathbb{R}\varepsilon{\mathbb{C}}\mathcal{O}$, 
	Institut d'Astrophysique de Paris, UMR 7095, CNRS, Sorbonne Universit{\'e}, 
	98\textsuperscript{bis} boulevard Arago, 75014 Paris, France}

\author{Achamveedu Gopakumar} \affiliation{Department of Astronomy and 
	Astrophysics, Tata Institute of Fundamental Research, Mumbai 400005, India}

\author{Bala R. Iyer}
\affiliation{International Centre for Theoretical Sciences, Tata Institute of 
	Fundamental Research, Bangalore 560089, India}

\begin{abstract}

  We compute the \emph{tail} contributions to the gravitational-wave mode
  amplitudes for compact binaries in eccentric orbits at the third
  post-Newtonian order of general relativity. We combine them with the already
  available instantaneous pieces and include the \emph{post-adiabatic} 
  corrections required to fully account for the effects of radiation-reaction 
  forces on the motion. We compare the resulting waveform in the small 
  eccentricity limit to the circular one, finding perfect agreement.

\end{abstract}


\maketitle

\section{Introduction}

In recent years the discoveries of gravitational waves (GW) by the LIGO and
Virgo Collaborations have opened a new window to the Universe~\cite{LIGO,
Virgo, GEO600, GW150914, GW151226, GW170104, GW170814, GW170817, GW170608,
GWTC-1}. KAGRA will join the global GW detector network in
2019~\cite{KAGRA} and LIGO-India in 2025~\cite{LIGO-India}, improving source 
localization and parameter estimation~\cite{lrr-prospects}, while LISA 
Pathfinder's exceptional performance~\cite{pathfinder-2016} -- showing that the 
LISA mission is feasible -- and maturing pulsar timing arrays~\cite{IPTA-2013} 
mark the beginning of multiwavelength, multiband GW astronomy.

Compact binary systems are the most prominent sources for the present and
future GW observatories. So far these events have been analyzed using
quasi-circular GW templates, as radiation-reaction effects tend to
circularize the orbits~\cite{peters-1963, peters-1964} for prototypical 
sources. For such systems one can thus assume that by the time the binary 
enters the sensitivity band of current ground-based detectors the eccentricity 
will be close to zero. However, there are a number of astrophysical scenarios 
in which binary systems could have moderate eccentricities when entering the 
sensitivity band of ground-based detectors~\cite{huerta-2016, tiwari-2016, 
gondan-2018-1, gondan-2018-2, rodriguez-2018, dorazio-2018, zevin-2018}. 
Recently, there have been studies showing that triple interactions among black 
holes can produce coalescing binaries with moderate eccentricities ($\sim 0.1$) 
when entering the LIGO band~\cite{samsing-2014, samsing-2017, samsing-2018-1} 
or large eccentricities ($\sim 0.9$) when entering the LISA 
band~\cite{bonetti-2018}. This has major implications on how to distinguish 
between binary black hole (BBH) formation channels~\cite{samsing-2018-2} and 
motivates the development of waveforms valid for nonzero eccentricities.

There has been great effort to model GWs of eccentric binary systems. One
usually employs the quasi-Keplerian parametrization~\cite{damour-1985, 
memmesheimer-2004} to describe the conservative binary orbits. The phasing
description, developed in Refs.~\cite{damour-2004, koenigsdoerffer-2006} and 
discussed in great detail for low-eccentricity binaries in 
Ref.~\cite{moore-2016}, efficiently incorporates the effects of radiation 
reaction, describing the binary dynamics on three different timescales: the 
orbital timescale, the periastron precession timescale, and the 
radiation-reaction timescale. In addition, the secular evolution of the orbital 
elements has been completed at the third post-Newtonian (3PN) order in 
Refs.~\cite{arun-2008-1, arun-2008-2, arun-2009}, including hereditary effects. 
Using this, several waveform models have been developed in the past 
years~\cite{yunes-2009, cornish-2010, key-2011, huerta-2014, huerta-2017, 
huerta-2018, gopakumar-2011, tanay-2016, hinder-2018, cao-2017, klein-2018}, 
for both nonspinning and spinning binaries.

In this paper, we extend the work in Ref.~\cite{mishra-2015} by computing the 
tail contributions to the GW amplitudes for compact binaries in eccentric 
orbits at the third post-Newtonian level. Combining our tail results with the
instantaneous ones, we then incorporate post-adiabatic 
corrections~\cite{damour-2004, koenigsdoerffer-2006, moore-2016} to get a
complete waveform including radiation-reaction effects valid during the early
inspiral of the binary system. We present all our results in modified harmonic
(MH) gauge in terms of the post-Newtonian parameter $\xb = ( G m \bar{\omega} / 
c^3 )^{2/3}$, where $G$ denotes the gravitational constant, $c$ the speed of 
light, $m$ the total mass of the binary, and $\bar{\omega}$ the adiabatic 
orbital frequency (see Sec.~\ref{sec: full waveform}), as well as a certain 
time eccentricity $\eb = \eb_t$ associated with the PN-accurate quasi-Keplerian 
parametrization. To calculate the complicated tail integrals, we work within a 
low-eccentricity expansion and express everything in terms of the mean anomaly 
$l$ and the phase angle $\lambda$, which accounts for the periastron advance. 
Compared to the results in Ref.~\cite{mishra-2015}, ours will thus not be valid 
for arbitrary eccentricities. Moreover, they will need to be completed by the 
memory contributions, which we will tackle in a follow-up 
paper~\cite{ebersold-2019}.

This paper is structured as follows: In Sec.~\ref{sec: prerequisites} we
quickly review the basics of spherical harmonic decomposition and recall how
to connect the radiative multipole moments to the actual source moments. We
also review the conservative 3PN-accurate quasi-Keplerian
parametrization~\cite{memmesheimer-2004}. In Sec.~\ref{sec: phasing}, we
discuss how to incorporate post-adiabatic corrections~\cite{damour-2004, 
koenigsdoerffer-2006} into this description. In Sec.~\ref{sec: hereditary}, we 
are then in a position to calculate the various tail integrals appearing in the 
source multipole moments. In Sec.~\ref{sec: full waveform}, we combine these 
results with the instantaneous ones and introduce post-adiabatic corrections. 
We also compare our results to the circular waveforms in 
Ref.~\cite{blanchet-2008}. Finally, in Sec.~\ref{sec: summary}, we give a brief 
summary of our work. Throughout this paper we mostly present results up to 
$\bigO(e)$, though expressions up to $\bigO(e^6)$ for all tail and 
post-adiabatic modes will be listed in a supplemental \emph{Mathematica} 
file~\cite{supplement}.

\section{Construction of the waveform for compact binaries in eccentric
  orbits}\label{sec: prerequisites}

\subsection{Polarizations and spherical-mode decomposition}

The gravitational waves emitted by an isolated system near future radiative
infinity are encoded in the transverse-traceless ($\TT$) projection 
$h_{ij}^\TT$ of the deviation of the space-time metric $g_{\mu\nu}$ from a flat 
metric $\eta_{\mu\nu}=\text{diag}(-1,1,1,1)$, in a radiative-type 
Cartesian-like coordinate grid $X^\mu = (cT, \bm{X})$, at order $1/R$, where $R 
= |\bm{X}|$ denotes the Euclidean distance of the vector $\bm{X}$ to the 
origin. It is convenient to chose this origin at the center of mass of the full 
system and to introduce the standard spherical coordinates $(\Theta, \Phi)$ 
associated with the so-defined Cartesian frame, for which the relation $X^i = R 
\, (\cos \Phi \sin \Theta, \sin \Phi \sin\Theta, \cos\Theta)$ holds. The 
radiative property of this frame ensures that a null geodesic going through the 
origin at time $T_R$ will reach an observer with position $\bm{X}$ at time 
$T=T_R + R/c$. If $\bm{N}(\Theta, \Phi)=\bm{X}/R$ denotes the unit direction of 
that observer, the plane span by the vectors $\bm{P}(\Theta, \Phi)$ and 
$\bm{Q}(\Theta, \Phi)$ belonging to some arbitrary direct orthonormal triad 
$(\bm{N},\bm{P},\bm{Q})$ must be transverse to the direction of propagation of 
wave rays.

The transverse-traceless projection $h_{ij}^\TT$ can be uniquely decomposed 
into symmetric trace-free (STF) radiative mass-type ($U_L$) and current-type 
($V_L$) multipole moments as:
\begin{align} \label{eq: hTT}
	h_{ij}^\TT =&\; \frac{4 G}{c^2 R} \PTT_{ijab}(\bm{N}) 
		\sum_{\ell=2}^{\infty} \frac{1}{c^\ell \ell!} \Big\{ N_{L-2} U_{ab L-2} 
		\nonumber\\
		&- \frac{2 \ell}{c (\ell + 1)} N_{c L-2}\epsilon_{cd(a} V_{b)d L-2} 
		\Big\} \Big|_{T_R} + \bigO \left( \frac{1}{R^2} \right) \,.
\end{align}
Here $\PTT_{ijab} = \PTT_{ia} \PTT_{jb} - \frac{1}{2} \PTT_{ij} \PTT_{ab}$,  
with $\PTT_{ij} = \delta_{ij} - N_i N_j$, is the $\TT$ projection operator. The 
waveform is usually projected on the transverse symmetric basis $e^+_{ij} = 
\frac{1}{2} (P_i P_j - Q_i Q_j)$, $e^\times_{ij} = P_{(i} Q_{j)}$,
\begin{align}
	\begin{pmatrix}
		h_+ \\
		h_\times
	\end{pmatrix}
	&= 
	\begin{pmatrix}
		e^+_{ij} \\
		e^\times_{ij}
	\end{pmatrix}
	\, h_{ij}^\TT \,,
\end{align}
the resulting components being referred to as the plus and cross polarizations, 
respectively. Equivalently the complex basis formed by the vector $\bm{m} = 
(\bm{P} + \ui \bm{Q}) / \sqrt{2}$ of spin weight 2 and its complex conjugate 
$\overline{\bm{m}}$ of spin weight $-2$ can be used. From the transverse 
trace-free character of the waveform, it follows that
\begin{align}
	h &= h_+ - \ui h_\times = h_{ij}^\TT \, \overline{m}^i \overline{m}^j \,.
\end{align}
From now on we shall assume that the vector $\bm{m}$ is proportional to
$\bm{m}_S = (\partial \bm{N} / \partial \theta + \ui \sin^{-1} \! \theta \, 
\partial \bm{N} / \partial \phi) / \sqrt{2}$ so that the functions adapted to 
the spherical decomposition of the spin $-2$ quantity $h$ are the usual 
spin-weighted spherical harmonics of weight $-2$, which will be denoted by 
$Y_{-2}^{\ell m}(\Theta, \Phi)$. In our conventions, they are given by
\begin{subequations}
\begin{align}
	Y_{-2}^{\ell m}(\Theta, \Phi) =&\; \sqrt{\frac{2\ell+1}{4\pi}} d_2^{\ell 
		m}(\Theta) e^{i m \Phi} \,,\\
	d_2^{\ell m} =&\; \sum_{k=k_\text{min}}^{k_\text{max}} \frac{(-1)^k}{k!} 
		\nonumber\\
		&\times \frac{\sqrt{(\ell+m)!(\ell-m)!(\ell+2)!(\ell-2)!}} {(k-m+2)! 
		(\ell+m-k)! (\ell-k-2)!} 
		\nonumber\\
		&\times \left(\cos\frac{\Theta}{2}\right)^{2\ell+m-2k-2} \left( \sin 
		\frac{\Theta}{2} \right)^{2k-m+2} \,,
\end{align}
\end{subequations}
with $k_\text{min} = \max(0,m-2)$ and $k_\text{max} = \min(\ell+m,\ell-2)$.
Thus, the gravitational waveform may be decomposed into spherical modes 
$h^{\ell m}$ as
\begin{align}\label{eq: mode decomposition}
	h_+ - \ui h_\times &= \sum_{\ell=2}^{+\infty} \sum_{m=-\ell}^{\ell} h^{\ell 
		m} Y_{-2}^{\ell m} (\Theta, \Phi) \,.
\end{align}
The spherical harmonic modes $h^{\ell m}$ can be written in terms of the
radiative mass-type ($U^{\ell m}$) and current-type ($V^{\ell m}$) multipole 
moments,
\begin{align}\label{eq: hlm rad mom}
	h^{\ell m} &= -\frac{G}{\sqrt{2}R c^{\ell+2}} \left(U^{\ell m} - 
		\frac{\ui}{c} V^{\ell m} \right) \,,
\end{align}
with the inverse relations
\begin{subequations}
\begin{align}
	U^{\ell m} &= -\frac{R c^{\ell +2}}{\sqrt{2}G} \left( h^{\ell m} + (-1)^m 
		\overline{h}{}^{\ell -m} \right) \,,\\
	V^{\ell m} &= -\frac{R c^{\ell + 3} }{\sqrt{2} \ui G} \left( -h^{\ell m} +
		(-1)^m \overline{h}{}^{\ell -m} \right) \,.
\end{align}
\end{subequations}
The radiative moments ($U^{\ell m}$, $V^{\ell m}$) are actually related to the 
STF radiative moments ($U_L$, $V_L$) by
\begin{subequations}\label{eq: radiative STF}
\begin{align}
	U^{\ell m} &= \frac{4}{\ell!} \sqrt{ \frac{(\ell+1) (\ell+2)}{2 \ell 
		(\ell-1)}} \alpha_L^{\ell m} U_L \,,\\
	V^{\ell m} &= -\frac{8}{\ell !} \sqrt{ \frac{\ell (\ell+2)}{2 (\ell+1) 
		(\ell-1)}} \alpha_L^{\ell m} V_L \,,
\end{align}
\end{subequations}
where the $\alpha_L^{\ell m}$ denote a set of constant STF tensors that connect 
the basis of spherical harmonics $Y^{\ell m}(\Theta, \Phi)$ to the set of STF 
tensors $N_{\la L \ra}$ as
\begin{subequations}
\begin{align}
	N_{\la L \ra}(\Theta, \Phi) &= \sum_{m=-\ell}^{\ell} \alpha_L^{\ell m} 
	Y^{\ell m} (\Theta, \Phi) \,,\\
	Y^{\ell m}(\Theta, \Phi) &= \frac{(2\ell+1)!!}{4\pi\ell!} 
		\overline{\alpha}_L^{\ell m} N^{\la L \ra}(\Theta, \Phi) \,.
\end{align}
\end{subequations}
They can be calculated through
\begin{align}
	\alpha_L^{\ell m} &= \int \ud\Omega\; N_{\la L \ra} \bar{Y}^{\ell m} \,,
\end{align}
and are given explicitly in Eq.~(2.12) of Ref.~\cite{thorne-1980}.

Remarkably, for planar binaries, there exists a mode 
separation~\cite{kidder-2008, faye-2012} such that $h^{\ell m}$ is completely 
determined by mass-type radiative multipole moments $U^{\ell m}$ for $\ell+m$ 
even and by current-type radiative multipole moments $V^{\ell m}$ for $\ell+m$ 
odd, hence
\begin{subequations}
\begin{align}
	h^{\ell m} &= -\frac{G}{\sqrt{2} R c^{\ell+2}} U^{\ell m} &&\textnormal{if 
		} \ell+m \textnormal{ is even} \,,\\
	h^{\ell m} &= \frac{\ui G}{\sqrt{2} R c^{\ell+3}} V^{\ell m} 
		&&\textnormal{if } \ell+m \textnormal{ is odd} \,.
\end{align}
\end{subequations}

Let us finally specify the choice of the Cartesian frame and polarization 
vectors in the case of interest where the source is a binary system of 
pointlike objects with bound orbits, since this choice will fully set the
amplitude modes computed in the present paper. We adopt the same conventions
as in Ref.~\cite{blanchet-2008}. In the absence of spin, the orbits stay in a
plane. The vector $\bm{e}_3$ is taken to be the unit normal vector orienting 
the sense of the motion positively. For the polarization vector $\bm{P}$, we 
pick the unit vector pointing towards the ascending node $\bm{N} \times 
\bm{e}_3$, with $\bm{N}$ representing the direction of the Earth observer. 
Therefore, we can also make it coincide with $\bm{e}_1$. To complete the
triads $\bm{e}_a$ and $(\bm{N},\bm{P},\bm{Q})$ we pose $\bm{e}_2=\bm{e}_3 \times
\bm{e}_1$ and $\bm{Q}=\bm{N}\times\bm{P}$. Notice that, by construction, 
$\bm{N}$ belongs to the plane spanned by $\{\bm{e}_2,\bm{e}_3\}$. Its spherical 
coordinates, in terms of the inclination of the binary $\iota$, are thus 
$(\Theta = \iota, \Phi = \pi/2)$.

\subsection{Multipole moments}\label{sec: multipole moments}

From Eqs.~(\ref{eq: hlm rad mom}--\ref{eq: radiative STF}), we see that we need
to relate the $U_L$ and $V_L$ to the actual source. In the multipolar 
post-Minkowsian (MPM) post-Newtonian (PN) formalism, the radiative moments 
($U_L$, $V_L$) are functionals of six sets of source moments ($I_L$, 
$J_L$, $W_L$, $X_L$, $Y_L$, $Z_L$). The relations between the radiative 
moments and the source moments have been obtained at the 3PN order and are 
listed in Ref.~\cite{blanchet-2008}, Eqs.~(5.4--5.11).

We can split the the expressions for the radiative moments into two parts,
namely the instantaneous and the hereditary parts:
\begin{align}
	U_L &= U_L^\inst + U_L^\hered \,.
\end{align}
The instantaneous contributions only depend on the state of the source at a
given retarded time, while the hereditary parts depend on, and thus require
knowledge of, the entire past history of the source. At leading order, the
instantaneous parts of the radiative moments are directly related to the
source moments as
\begin{subequations}
\begin{align}
	U_L^\inst(t_r) &= I_L^{(\ell)}(t_r) + \bigO(c^{-3}) \,,\\
	V_L^\inst(t_r) &= J_L^{(\ell)}(t_r) + \bigO(c^{-3}) \,, 
\end{align}
\end{subequations}
with $t_r$ denoting here a ``dummy'' variable. Corrections from the gauge
moments ($W_L$, $X_L$, $Y_L$, $Z_L$) enter at higher orders. In this work, we 
will focus on the hereditary tail contributions. For a complete treatment of 
the instantaneous contributions, we refer to Ref.~\cite{mishra-2015}.

To the desired accuracy, the hereditary contributions to the radiative moments
are given by
\begin{widetext}
\begin{subequations}\label{eq: U_L}
\begin{align}
	\label{eq: U_ij}
	U_{ij}^\hered (t_r) =&\; \frac{2GM}{c^3} \int_{0}^{\infty} \ud\tau\; \left[ 
		\ln\left( \frac{\tau}{2\tau_0} \right) + \frac{11}{12} \right] 
		I_{ij}^{(4)}(t_r - \tau) - \frac{2G}{7c^5} \int_{-\infty}^{t_r} 
		\ud\tau\; I_{a\la i}^{(3)}(\tau) I_{j\ra a}^{(3)}(\tau) \nonumber\\
		&+ 2 \left( \frac{GM}{c^3} \right)^2 \int_{0}^{\infty} \ud\tau\; \left[ 
		\ln^2\left( \frac{\tau}{2\tau_0} \right) + \frac{57}{70} \ln\left( 
		\frac{\tau}{2\tau_0} \right) + \frac{124627}{44100} \right] 
		I_{ij}^{(5)}(t_r - \tau) + \bigO(c^{-7}) \,,\\
	\label{eq: U_ijk}
	U_{ijk}^\hered (t_r) =&\; \frac{2GM}{c^3} \int_{0}^{\infty} \ud\tau\; 
		\left[ \ln\left( \frac{\tau}{2\tau_0} \right) + \frac{97}{60} \right] 
		I_{ijk}^{(5)}(t_r - \tau) \nonumber\\
		&+ \frac{G}{c^5} \int_{-\infty}^{t_r} \ud\tau\; \left[ -\frac{1}{3} 
		I_{a\la i}^{(3)}(\tau) I_{jk\ra a}^{(4)}(\tau) - \frac{4}{5} 
		\epsilon_{ab\la i} I_{ja}^{(3)}(\tau) J_{k\ra b}^{(3)}(\tau) \right] + 
		\bigO(c^{-6}) \,,\\
	\label{eq: U_ijkl}
	U_{ijkl}^\hered (t_r) =&\; \frac{2GM}{c^3} \int_{0}^{\infty} \ud\tau\; 
		\left[ \ln\left( \frac{\tau}{2\tau_0} \right) + \frac{59}{30} \right] 
		I_{ijkl}^{(6)}(t_r - \tau) + \frac{2G}{5c^3} \int_{-\infty}^{t_r} 
		\ud\tau\; I_{\la ij}^{(3)}(\tau) I_{kl\ra}^{(3)}(\tau) + \bigO(c^{-5}) 
		\,,\\
	\label{eq: U_ijklm}
	U_{ijklm}^\hered (t_r) =&\; \frac{2GM}{c^3} \int_{0}^{\infty} \ud\tau\; 
		\left[ \ln\left( \frac{\tau}{2\tau_0} \right) + \frac{232}{105} \right] 
		I_{ijklm}^{(7)}(t_r - \tau) + \frac{20G}{21c^3} \int_{-\infty}^{t_r} 
		\ud\tau\; I_{\la ij}^{(3)}(\tau) I_{klm\ra}^{(4)}(\tau) + \bigO(c^{-4}) 
		\,,
\end{align}
\end{subequations}
\begin{subequations}\label{eq: V_L}
\begin{align}
	\label{eq: V_ij}
	V_{ij}^\hered (t_r) =&\; \frac{2GM}{c^3} \int_{0}^{\infty} \ud\tau\; \left[ 
		\ln\left( \frac{\tau}{2\tau_0} \right) + \frac{7}{6} \right] 
		J_{ij}^{(4)}(t_r - \tau) + \bigO(c^{-6}) \,,\\
	\label{eq: V_ijk}
	V_{ijk}^\hered (t_r) =&\; \frac{2GM}{c^3} \int_{0}^{\infty} \ud\tau\; 
		\left[ \ln\left( \frac{\tau}{2\tau_0} \right) + \frac{5}{3} \right] 
		J_{ijk}^{(5)}(t_r - \tau) + \bigO(c^{-5}) \,,\\
	\label{eq: V_ijkl}
	V_{ijkl}^\textnormal{hered} (t_r) =&\; \frac{2GM}{c^3} \int_{0}^{\infty} 
		\ud\tau\; \left[ \ln\left( \frac{\tau}{2\tau_0} \right) + 
		\frac{119}{60} \right] J_{ijkl}^{(6)}(t_r - \tau) + \bigO(c^{-4}) \,,
\end{align}
\end{subequations}
\end{widetext}
where $M = m (1 - \nu x / 2)+\bigO(c^{-4})$ is the Arnowitt-Deser-Misner (ADM)
mass of the source, $m = m_1 + m_2$ the total mass, $\nu = m_1 m_2 / m^2$ the
symmetric mass ratio, and $\tau_0$ an arbitrary length scale originally
introduced in the MPM formalism. None of the other moments contributes to the
hereditary part of the waveform~(\ref{eq: hTT}) at 3PN order, since
\begin{subequations}
\begin{align}
	U_{L>5}^\hered &= \bigO(c^{-3}) \,, \\
	V_{L>4}^\hered &= \bigO(c^{-3}) \,.
\end{align}
\end{subequations}

In the above hereditary contributions, there are two different types of
integrals: those with logarithms and those without. The logarithmic integral 
in the first line of Eq.~(\ref{eq: U_ij}) is called the tail integral while the 
one on the second line is the tails-of-tails integral. On the other hand, the 
integral without a logarithmic kernel is the memory integral. Note that there 
are no memory contributions to the radiative current moments $V_L$. Physically, 
wave tails come from the scattering of the linear waves, generated by the 
matter source, off the space-time curvature due to the total ADM mass of the 
isolated system. It is a (power of) monopole-wave interaction effect with a 
weak past dependence. By contrast, the memory pieces of the waves are produced 
by the effective stress-energy tensor of the source radiation itself. It is a 
wave-wave interaction effect with a strong past dependence~\cite{blanchet-1992}.

The expressions for the source moments ($I_L$, $J_L$) in terms of the binary
separation $r$, its time derivative $\dot{r}$, the polar angle $\phi$ of the
relative position, and its derivative $\dot{\phi}$ are now required. Observing 
Eqs.~(\ref{eq: U_L}--\ref{eq: V_L}), we note that $I_{ij}$, $J_{ij}$ and 
$I_{ijk}$ are needed to an accuracy of 1PN, while all other multipole moments 
are only needed to leading Newtonian order. The relevant expressions are listed 
in Ref.~\cite{arun-2008-2} using standard harmonic (SH) coordinates. The 
logarithms appearing at 3PN order in the SH gauge can, however, be transformed 
away in appropriate modified harmonic coordinates, as demonstrated Sec.~IV~B 
of Ref.~\cite{arun-2008-2}. For the hereditary parts, this will not make any
difference, as we shall only need relative 1PN-accurate expressions for
certain ($I_L$, $J_L$), but, when adding up instantaneous terms from 
Ref.~\cite{mishra-2015} to our hereditary parts, we shall always work within
the MH gauge. The binary separation vector will be represented by
$x^i\equiv r\, n^i$, whereas $v^i=\ud x^i/\ud t$ will stand for the relative
velocity. The expressions relevant for the calculation of the hereditary parts
are
\begin{widetext}
\begin{subequations}\label{eq: I_L}
\begin{align}
	\label{eq: I_ij}
	I_{ij} =&\; \nu m \left( A_1\, x_{\la ij\ra} + A_2\, \frac{r \dot{r}}{c^2} 
		x_{\la i} v_{j\ra} + A_3\, \frac{r^2}{c^2} v_{\la ij\ra}\right) + 
		\bigO(c^{-7}) \,,\\
	\label{eq: I_ijk}
	I_{ijk} =&\; -\nu m \Delta \left( B_1\, x_{\la ijk\ra} + B_2\, \frac{r 
		\dot{r}}{c^2} x_{\la ij} v_{j\ra} + B_3\, \frac{r^2}{c^2} x_{\la i} 
		v_{jk\ra}\right) + \bigO(c^{-6}) \,,\\
	\label{eq: I_ijkl}
	I_{ijkl} =&\; \nu m (1-3\nu) x_{\la ijkl\ra} + \bigO(c^{-5}) \,,\\
	\label{eq: I_ijklm}
	I_{ijklm} =&\; -\nu m \Delta (1-2\nu) x_{\la ijklm\ra} + \bigO(c^{-4}) \,,
\end{align}
\end{subequations}
\begin{subequations}\label{eq: J_L}
\begin{align}
	\label{eq: J_ij}
	J_{ij} =&\; -\nu m \Delta \left( C_1\, \epsilon_{ab\la i} x_{j\ra a} v_b + 
		C_2\, \frac{r \dot{r}}{c^2} \epsilon_{ab\la i} v_{j\ra b} x_{a} \right) 
		+ \bigO(c^{-6}) \,,\\
	\label{eq: J_ijk}
	J_{ijk} =&\; \nu m (1-3\nu) \epsilon_{ab\la i} x_{jk\ra a} v_b + 
		\bigO(c^{-5}) \,,\\
	\label{eq: J_ijkl}
	J_{ijkl} =&\; -\nu m \Delta (1-2\nu) \epsilon_{ab\la i} x_{jkl\ra a} v_b + 
		\bigO(c^{-4}) \,,
\end{align}
\end{subequations}
where $\Delta = (m_1 - m_2) / m$ is the mass difference ratio and the constants 
$A_i$, $B_i$, and $C_i$ read
\begin{subequations}
\begin{align}
	A_1 =&\; 1 + \frac{1}{c^2} \left[ v^2 \left( \frac{29}{42} - \frac{29 
		\nu}{14} \right) + \frac{Gm}{r} \left( -\frac{5}{7} + \frac{8\nu}{7} 
		\right) \right] \,,\\
	A_2 =&\; -\frac{4}{7} + \frac{12\nu}{7} \,,\\
	A_3 =&\; \frac{11}{21} - \frac{11\nu}{7} \,,\\
	B_1 =&\; 1 + \frac{1}{c^2} \left[ v^2 \left( \frac{5}{6} - \frac{19\nu}{6} 
		\right) + \frac{Gm}{r} \left( -\frac{5}{6} + \frac{13\nu}{6} \right) 
		\right] \,,\\
	B_2 =&\; -(1-2\nu) \,,\\
	B_3 =&\; 1-2\nu \,,\\
	C_1 =&\; 1 + \frac{1}{c^2} \left[ v^2 \left( \frac{13}{28} - \frac{17 
		\nu}{7} \right) + \frac{Gm}{r} \left( \frac{27}{14} + \frac{15\nu}{7} 
		\right) \right] \,,\\
	C_2 =&\; \frac{5}{28} (1-2\nu) \,.
\end{align}
\end{subequations}
\end{widetext}

\subsection{Quasi-Keplerian parametrization}\label{sec: keplerian 
parametrization}

The expressions in Eqs.~(\ref{eq: I_L}--\ref{eq: J_L}) in terms of the
variables ($r$, $\dot{r}$, $\phi$, $\dot{\phi}$) are the most general ones.
Now, when calculating the tail integrals, we should replace the latter
quantities by their actual analytic time evolution for eccentric orbits. At
the third post-Newtonian order, the conservative orbital dynamics of compact
binaries in eccentric orbits is specified by providing the following
generalized quasi-Keplerian parametrization~\cite{memmesheimer-2004} for the
dynamical variables $r$ and $\phi$:
\begin{subequations}\label{eq: quasi-keplerian}
\begin{align}
	r =\;& a_r \left(1 - e_r \cos u \right)	\,,\\
	\phi - \phi_{0} =\;& (1 + k ) v + \left(f_{4\phi} + f_{6\phi} \right) \sin 
		(2v) \nonumber\\
		&+ \left(g_{4\phi} + g_{6\phi} \right) \sin (3v) + i_{6\phi}\sin (4v) 
		\nonumber\\
		&+ h_{6\phi} \sin (5v) \label{eq: quasi-keplerian phi} \,,\\
	\text{where} \quad 
	v =&\; 2 \arctan \left[\left( \frac{ 1 + e_{\phi} }{1 - e_{\phi}} 
		\right)^{1/2} \tan \frac{u}{2} \right] \,.
\end{align}
\end{subequations}
An interesting feature in the above equations is the presence of different
eccentricity parameters $e_r$ and $e_\phi$, introduced in such a way that the
parametrization looks ``Keplerian''. The parameter $k$ is nothing but the 
periastron advance per orbital revolution. The parameters $a_r$, $e_r$, and 
$e_\phi$ are the PN-accurate semi-major axis and the radial and angular 
eccentricities, while $f_{4\phi}$, $f_{6\phi}$, $g_{4\phi}$, $g_{6\phi}$, 
$i_{6\phi}$, and $h_{6\phi}$ are some orbital functions of the energy and 
angular momentum that enter at the 2PN and 3PN orders. The explicit expressions 
are available in Ref.~\cite{memmesheimer-2004}.

The eccentric anomaly $u$ is linked to the mean anomaly $l$ through the 
3PN-accurate Kepler equation
\begin{align}\label{eq: 3PN_KE}
	l =&\; u - e_t \sin u + \left(g_{4t} + g_{6t} \right)(v-u) \nonumber\\
		&+ \left(f_{4t} + f_{6t} \right)\sin v + i_{6t} \sin (2v) + h_{6t} \sin 
		(3v)\,.
\end{align}
Here, $e_t$ is another eccentricity parameter, usually called the time
eccentricity, and the functions $g_{4t}$, $g_{6t}$, $f_{4t}$, $f_{6t}$, 
$i_{6t}$, and $h_{6t}$ are additional 2PN and 3PN orbital functions of the 
energy and angular momentum. Together, Eqs.~(\ref{eq: quasi-keplerian}) and 
(\ref{eq: 3PN_KE}) fully parametrize the conservative orbital dynamics of 
compact binaries on eccentric orbits. Note that we choose to express all our 
equations in terms of the post-Newtonian parameter $x = (Gm\omega / c^3)^{2/3}$ 
and the time eccentricity $e = e_t$, with $\omega = (1+k)n$ being the orbital 
frequency and $n = 2 \pi / P$ the mean motion associated with the period $P$. 
In the next section, we shall introduce post-adiabatic corrections to this 
quasi-Keplerian description. We will then have to replace the parameters ($x$, 
$e$) with their slowly evolving counterparts ($\xb$, $\eb$).

The appearance of the periastron precession at first post-Newtonian order
introduces a double periodic motion on two timescales: the orbital timescale 
and the precession timescale. It is thus customary to split the phase $\phi$
into an angle $\lambda$ that is linear in $l$ and an oscillatory part $W(l)$
that is $2\pi$-periodic in $l$~\cite{gopakumar-2002, damour-2004, 
tessmer-2007}. This leads us to write
\begin{subequations}
\begin{align}
	\phi =\;& \lambda + W(l) \,,\\
	\lambda =\;& \phi_0 + (1+k)l \,,\\
	W(l) =\;& (1+k)(v-l) + \left(f_{4\phi} + f_{6\phi} \right) \sin (2v) 
		\nonumber\\
		&+ \left(g_{4\phi} + g_{6\phi} \right) \sin (3v) + i_{6\phi}\sin (4v) 
		\nonumber\\
		&+ h_{6\phi} \sin (5v) \label{eq: quasi-keplerian W}\,,
\end{align}
\end{subequations}
with $\phi_0$ denoting the initial polar angle at $u=0$.

To evaluate the various time integrals appearing in the tail contributions to 
the waveform, we will need explicit expressions for $u$ and $\phi$ in terms of
the angles $l$ and $\lambda$. This can be achieved by solving the Kepler 
equation (\ref{eq: 3PN_KE}). We employ the method described in 
Ref.~\cite{boetzel-2017}, which yields
\begin{subequations}\label{eq: KE solution}
\begin{align}
	u &= l + \sum_{s=1}^{\infty} A_s \sin(sl) \,,\\
	A_s &= \frac{2}{s} J_s(s e_t) + \sum_{j=1}^{\infty} \alpha_j \left\{ 
		J_{s+j}(s e_t) - J_{s-j}(s e_t) \right\} \,,
\end{align}
\end{subequations}
where the constants $\alpha_j$ are some PN-accurate functions of the energy and 
angular momentum entering at the second post-Newtonian order. It remains to 
display an explicit expression for the $2\pi$-periodic function $W(l)$ in terms 
of $l$,
\begin{subequations}\label{eq: W solution}
\begin{align}
	W(l) =\;& \sum_{s=1}^{\infty} \mathcal{W}_s \sin(sl) \,,\\
	\mathcal{W}_s =\;& (1+k) B_s + \left(f_{4\phi} + f_{6\phi} \right) 
		\sigma_s^{2v} \nonumber\\
		&+ \left(g_{4\phi} + g_{6\phi} \right) \sigma_s^{3v} + i_{6\phi} 
			\sigma_s^{4v} + h_{6\phi} \sigma_s^{5v} \,,
\end{align}
\end{subequations}
with the constants $B_s$ and $\sigma_s^{jv}$ given in Eqs.~(C8) and (32b) of 
Ref.~\cite{boetzel-2017}. We finally find, expanding to $\bigO(x^3)$ and
$\bigO(e)$,
\begin{subequations}\label{eq: KE solution expanded}
\begin{align}
	\label{eq: u solution}
	u =&\; l + e \sin(l) + x^2 \left(-\frac{15}{2} + \frac{9\nu}{8} + 
		\frac{\nu^2}{8} \right) e\sin(l) \nonumber\\
		&+ x^3 \left( -55 + \frac{104593\nu}{1680} + \frac{3\nu^2}{4} + 
		\frac{\nu^3}{24} \right) e \sin(l) \,,\\
	\label{eq: phi solution}
	\phi =&\; \lambda + 2 e \sin(l) + x (10-\nu) e\sin(l) \nonumber\\
		&+ x^2 \left( 52 - \frac{235\nu}{12} + \frac{\nu^2}{12} \right) e 
		\sin(l) \nonumber\\
		&+ x^3 \bigg( 292 + \left(-\frac{420131}{840} + \frac{287\pi^2}{32} 
		\right) \nu \nonumber\\
		&+ \frac{521\nu^2}{24} + \frac{\nu^3}{24} \bigg) e\sin(l) \,.
\end{align}
\end{subequations}
We shall use these expressions to write the source multipole moments ($I_L$,
$J_L$) in terms of $l$ and $\lambda$.

\section{Phasing of the orbital elements}\label{sec: phasing}

So far, we used the conservative quasi-Keplerian description of the dynamics of
nonspinning compact binaries. This analytic parametrization is possible due to
the fact that the conservative problem admits four integrals of motion, or
even two, when the problem is restricted to the orbital plane. In our case, 
those two integrals are encoded in the two intrinsic constants $x$ and $e=e_t$. 
There also exist two extrinsic constants $c_l$ and $c_\lambda$,
\begin{subequations}
\begin{align}
	l(t) &= n(t-t_0) + c_l \,, \\
	\lambda(t) &= (1+k) n (t-t_0) + c_\lambda \,,
\end{align}
\end{subequations}
corresponding to the initial values of the two phase angles $l$ and $\lambda$,
respectively. We now move to include phasing effects due to energy and angular
momentum loss into this quasi-Keplerian parametrization. An efficient
description of the dynamics of nonspinning compact binaries with phasing
is presented in Refs.~\cite{damour-2004,koenigsdoerffer-2006}. Following 
Ref.~\cite{damour-1983}, they employ a method of \emph{variation of constants} 
where the constants of motion of the conservative problem ($x$, $e$, $c_l$, 
$c_\lambda$) are treated as time-varying quantities. Specifically, the 
post-Newtonian parameter $x = x(t)$ and the time eccentricity $e = e(t)$ are 
now genuine functions of time, while the angles $l$ and $\lambda$ are given by
\begin{subequations}
\begin{align}
	l(t) &= \int_{t_0}^{t} n(t') \ud t' + c_l(t) \,, \\
	\lambda(t) &= \int_{t_0}^{t} [1+k(t')] n(t') \ud t' + c_\lambda(t) \,.
\end{align}
\end{subequations}

To obtain the evolution of the functions $c_\alpha(t) = (x(t), e(t), c_l(t), 
c_\lambda(t))$, one starts from the PN-accurate equations of motion
\begin{subequations}
\begin{align}
	\dot{\bf{x}} &= \bf{v} \,, \\
	\dot{\bf{v}} &= \mathcal{A}_0(\bf{x}, \bf{v}) + \mathcal{A}'(\bf{x}, 
		\bf{v}) \,,
\end{align}
\end{subequations}
with $\mathcal{A}_0$ being the conservative and $\mathcal{A}'$ the dissipative
piece of the equations of motion. These equations are first solved neglecting 
the dissipative term $\mathcal{A}'$, leading to the conservative 
quasi-Keplerian description of Sec.~\ref{sec: keplerian parametrization}. The
full solution including radiation reaction is then found by varying the 
``constants'' $c_\alpha(t)$, leading to differential equations of the form
\begin{align}
	\frac{\ud c_\alpha}{\ud l} &= G_\alpha(l, c_\alpha) \,.
\end{align}
One can then introduce a two-scale decomposition of all phase variables
$c_\alpha(l)$ into a slow (radiation-reaction timescale) secular drift and a
fast (orbital timescale) periodic oscillation as
\begin{align}
	c_\alpha(t) = \bar{c}_\alpha(t) + \tilde{c}_\alpha(t) \,,
\end{align}
with
\begin{subequations}
\begin{align}
	\frac{\ud\bar{c}_\alpha}{\ud l} &= \bar{G}_\alpha(l, c_\alpha) \label{eq: 
		secular evolution} \,,\\
	\frac{\ud\tilde{c}_\alpha}{\ud l} &= \tilde{G}_\alpha(l, c_\alpha) = 
		G_\alpha(l, c_\alpha) - \bar{G}_\alpha(l, c_\alpha) \label{eq: osc 
		evolution} \,,
\end{align}
\end{subequations}
$\bar{G}_\alpha$ and $\tilde{G}_\alpha$ here being the orbital averaged and
oscillatory pieces of $G_\alpha$. The secular evolution of the orbital
elements (\ref{eq: secular evolution}) can also be derived from the heuristic
balance equations $\la \ud E/\ud t \ra = - \la \mathcal{F} \ra$ and
$\la \ud J/\ud t \ra = - \la \mathcal{G} \ra$, where $\mathcal{F}$ is the 
energy flux and $\mathcal{G}$ the angular momentum flux. This approach is 
discussed at the 3PN order in a series of papers~\cite{arun-2008-1, 
arun-2008-2, arun-2009}, which notably take care of the hereditary 
contributions to the energy and angular momentum fluxes.

After the above procedure is applied, we have
\begin{subequations}\label{eq: two-scale decomp}
\begin{align}
	x(t) &= \xb(t) + \xp(t) \,, \\
	e(t) &= \eb(t) + \ep(t) \,, \\
	c_l(t) &= \clb + \clp(t) \,, \\
	c_\lambda(t) &= \clamb + \clamp(t) \,,
\end{align}
\end{subequations}
where $\clb$ and $\clamb$ are found to be true integration constants. The
secular evolution of the orbital elements $\bar{n}(t)$, $\bar{k}(t)$, $\xb(t)$,
and $\eb(t)$ is given in Sec.~VI of Ref.~\cite{arun-2009}. At leading order, 
these equations reduce to the famous formulas by Peters and 
Mathews~\cite{peters-1963,peters-1964}:
\begin{subequations}\label{eq: peters-mathews}
\begin{align}
	\frac{\ud\xb}{\ud t} &= \frac{c^3 \nu}{Gm} \frac{\xb^5}{(1-\eb^2)^{7/2}} 
		\left( \frac{64}{5} + \frac{584}{15} \eb^2 + \frac{74}{15} \eb^4 
		\right) \,,\\
	\frac{\ud\eb}{\ud t} &= -\frac{c^3 \nu}{Gm} \frac{\eb \, 
		\xb^4}{(1-\eb^2)^{5/2}} \left( \frac{304}{15} + \frac{121}{15} \eb^2 
		\right) \,.
\end{align}
\end{subequations}

The periodic variations in Eqs.~(\ref{eq: two-scale decomp}) can be computed
from Eqs.~(34) and (35) of Ref.~\cite{koenigsdoerffer-2006} and are explicitly 
given in Eqs.~(36). Note, though, that there is an error in the expressions for
$\clp$ and $\clamp$ provided by Eqs.~(36c) and (36d) of that paper. Indeed, the
periodic variations $\clp$ and $\clamp$ refer to the zero-average oscillatory
contributions to $c_l$ and $c_\lambda$. They are found by integrating Eqs.~(35) 
and then subtracting the orbital average, i.e., finding the unique zero-average 
primitive, so that we are left with a purely oscillatory solution. Now, we find 
that, unfortunately, the explicit orbital averages of Eqs.~(36c) and (36d) in 
Ref.~\cite{koenigsdoerffer-2006} do not give zero. This is because the 
averaging of these terms is performed over the eccentric anomaly $\ud u$, 
whereas the orbital averaging requires integrating temporal variations over an 
orbital period and, therefore, should be done using $\ud l = (1 - e \cos u) \ud 
u$. We show below the corrected expressions for $\clp$ and $\clamp$ in terms of 
$e_t = \eb$, $\xi = \xb^{3/2}$ and $u = \bar{u}$, as they appear in 
Ref.~\cite{koenigsdoerffer-2006}:
\begin{widetext}
\begin{subequations}\label{eq: corrected cl}
\begin{align}
	\clp =\;& -\frac{2 \xi^{5/3} \nu}{45 e_t^2} \bigg\{ \frac{144 e_t^2}{\chi} 
		+ \frac{18 - 258 e_t^2}{\chi^2} + \frac{-56 + 92 e_t^2 - 36 
		e_t^4}{\chi^3} + \frac{105 (1 -	e_t^2)^2}{\chi^4} \nonumber\\
		&- \frac{1}{2 (1 - e_t^2)^{1/2}} \left[ 134 - 339 e_t^2 + 288 e_t^2 
		\sqrt{1 - e_t^2} \right] \bigg\} + \bigO(\xi^{7/3}) \,,\\
	\clamp =\;& \frac{2 \xi^{5/3} \nu}{45 e_t^2} \bigg\{ \left[ 
		\frac{18}{\chi^2} - \frac{56 - 36 e_t^2}{\chi^3} + \frac{105 (1 - 
		e_t^2)}{\chi^4} \right] \sqrt{1 - e_t^2} - \frac{144 e_t^2}{\chi} - 
		\frac{18 - 258 e_t^2}{\chi^2} + \frac{56 - 92 e_t^2 + 36 e_t^4}{\chi^3} 
		- \frac{105 (1 - e_t^2)^2}{\chi^4} \nonumber\\
		&- \frac{1}{2 (1 - e_t^2)} \left[ 134 - 147 e_t^2 + 288 e_t^4 - \left( 
		134 - 339 e_t^2 \right) \sqrt{1 - e_t^2} \right] \bigg\} + 
		\bigO(\xi^{7/3}) \,.
\end{align}
\end{subequations}
\end{widetext}

Similarly, we split the angles $l$ and $\lambda$ into orbital averaged and
oscillatory contributions
\begin{subequations}\label{eq: two-scale decomp 2}
\begin{align}
	l(t) &= \lb(t) + \lp(t) \,, \\
	\lambda(t) &= \lab(t) + \lap(t) \,,
\end{align}
\end{subequations}
with $\lb(t)$ and $\lab(t)$ defined by
\begin{subequations}\label{eq: secular l and la}
\begin{align}
	\lb(t) &= \int_{t_0}^{t} \bar{n}(t')\ud t' + \clb \,,\\
	\lab(t) &= \int_{t_0}^{t} [1+\bar{k}(t')] \bar{n}(t') \ud t' + \clamb \,.
\end{align}
\end{subequations}
The oscillatory contributions $\lp$ and $\lap$ are calculated as in Eqs.~(39) 
of Ref.~\cite{koenigsdoerffer-2006},
\begin{subequations}
\begin{align}
	\lp(\lb) &= \int \frac{\tilde{n}}{\bar{n}} \ud l + \clp(\lb) \,, \\
	\lap(\lb) &= \int \left[ (1 + \bar{k}) \frac{\tilde{n}}{\bar{n}} + 
		\tilde{k} \right] \ud l + \clamp(\lb) \,,
\end{align}
\end{subequations}
where $\tilde{k} = (\partial k / \partial n) \tilde{n} + (\partial k / \partial 
e_t) \ep_t$ denotes the periodic part of $k$ and the integrals again mean the 
unique zero-average primitives. Equations~(40) for $\lp$ and $\lap$  in 
Ref.~\cite{koenigsdoerffer-2006} are erroneous, since they do not 
average to zero either. We list below the corrected expressions:
\begin{widetext}
\begin{subequations}\label{eq: corrected lp}
\begin{align}
	\lp(l) =\;& \frac{\xi^{5/3} \nu}{15 (1 - e_t^2)^3} \bigg\{ (602 + 673 
		e_t^2) \chi + (314 - 203 e_t^2 - 111 e_t^4) \ln \chi - (602 + 673 
		e_t^2) + \frac{-98 + 124 e_t^2 + 46 e_t^4 - 72 e_t^6}{\chi} \nonumber\\
		&- \frac{105 (1 - e_t^2)^3}{\chi^2} - \frac{1}{2} \bigg[ 432 + 444 
		e_t^2 + 543 e_t^4 - 144 e_t^6 - (838 - 826 e_t^2 - 12 e_t^4) \sqrt{1 - 
		e_t^2} + (628 - 406 e_t^2 - 222 e_t^4) \nonumber\\
		&\times \ln \bigg( \frac{1 + \sqrt{1 - e_t^2}}{2} \bigg) \bigg] \bigg\} 
		+ \frac{\xi^{5/3} \nu}{5 (1 - e_t^2)^{7/2}} \left( 96 + 292 e_t^2 + 37 
		e_t^4 \right) \int \left[ 2 \tan^{-1} \left( \frac{\beta_t \sin u}{1 - 
		\beta_t \cos u} \right) + e_t \sin u \right] \chi \ud u \nonumber\\
		&+ \clp(l) + \bigO(\xi^{7/3}) \,,\\
	\lap(l) =\;& \lp(l) - \clp(l) + \clamp(l) + \bigO(\xi^{7/3}) \,.
\end{align}
\end{subequations}
\end{widetext}
The errors in Eqs.~(36c), (36d), and (40) of Ref.~\cite{koenigsdoerffer-2006}, 
though, do not affect the other equations of that work. We refer to 
Appendix~\ref{sec: integrals} for some integral relations necessary to compute 
the zero-average primitives.

We finally give expressions for the oscillatory contributions $\xp$, $\ep$,
$\lp$, and $\lap$ in terms of the slowly evolving variables $\xb$, $\eb$, and
$\lb$. We list here the expressions to $\bigO(\eb^2)$:
\begin{subequations}\label{eq: periodic variations}
\begin{align}
	\xp(t) =\;& \nu \xb^{7/2} \eb \bigg[ 80 \sin(\lb) + \frac{1436}{15} \eb 
		\sin(2\lb) \nonumber\\
		&+ \eb^2 \left( \frac{4538}{15} \sin(\lb) + \frac{6022}{45} \sin(3\lb) 
		\right) \bigg] \nonumber\\
		&+ \bigO(\xb^{9/2}) \,, \\
	\ep(t) =\;& -\nu \xb^{5/2} \bigg[ \frac{64}{5} \sin(\lb) + \frac{352}{15} 
		\eb \sin(2\lb) \nonumber\\
		&+ \eb^2 \left( \frac{1138}{15} \sin(\lb) + \frac{358}{9} \sin(3\lb) 
		\right) \bigg] \nonumber\\
		&+ \bigO(\xb^{7/2}) \,,\\
	\lp(t) =\;& -\nu \xb^{5/2} \bigg[ \frac{64}{5\eb} \cos(\lb) + 
		\frac{352}{15} \cos(2\lb) \nonumber\\
		&+ \eb \left( \frac{1654}{15} \cos(\lb) + \frac{358}{9} \cos(3\lb) 
		\right) \nonumber\\ 
		&+ \eb^2 \left( \frac{694}{15} \cos(2\lb) + \frac{1289}{20} \cos(4\lb) 
		\right) \bigg] \nonumber\\
		&+ \bigO(\xb^{7/2}) \,,\\
	\lap(t) =\;& -\nu \xb^{5/2} \bigg[ \frac{296}{3} \eb \cos(\lb) + 
		\frac{199}{5} \eb^2 \cos(2\lb) \bigg] \nonumber\\
		&+ \bigO(\xb^{7/2}) \,.
\end{align}
\end{subequations}
These results agree with Eqs.~(4.9) of Ref.~\cite{moore-2016}, except two 
constant terms in $\lp(t)$ and $\lap(t)$, due to the already mentioned incorrect
average. Indeed, all our results are purely oscillatory, zero-average functions
and thus correctly describe the periodic post-adiabatic corrections.

Given the waveform in terms of the conservative quasi-Keplerian
parametrization, one can then include post-adiabatic effects by making the
simple substitutions
\begin{subequations}\label{eq: phasing subst}
\begin{align}
	x &\rightarrow \xb + \xp \,, \\
	e &\rightarrow \eb + \ep \,, \\
	l &\rightarrow \lb + \lp \,, \\
	\lambda &\rightarrow \lab + \lap \,.
\end{align}
\end{subequations}
As all of the periodic (tilde) contributions are of relative 2.5PN order
compared to the slowly evolving (bar) parts, we only have to make these
substitutions at leading Newtonian and 0.5PN order in the $h^{\ell m}$ to be
accurate to 3PN order. In all higher-order terms, we can simply replace the
variables ($x$, $e$, $l$, $\lambda$) by their secular evolving parts ($\xb$,
$\eb$, $\lb$, $\lab$).

Note that Eq.~(\ref{eq: quasi-keplerian phi}) gives the relation between the
geometrical phase $\phi$ and the angles $l$ and $\lambda$. We can rewrite this
relation in terms of the slowly evolving angles $\lb$ and $\lab$ and find
\begin{align}\label{eq: quasi-keplerian phi bar}
	\phi =\;& \lambda + W(l) = \lab + \bar{W}(\lb) + \lap + (\vp - \lp) \,,
\end{align}
where $\bar{W}(\lb)$ is given by Eq.~(\ref{eq: quasi-keplerian W}), but with
all quantities on the RHS replaced with their secular evolving parts, and the
periodic variation $\vp$ of the true anomaly is given by
\begin{align}
	\vp &= \frac{\pd \vb}{\pd \ub} \, \up + \frac{\pd \vb}{\pd \eb} \, \ep 
		\nonumber\\
		&= \frac{\sqrt{1 - \eb^2}}{1 - \eb \cos \ub} \up + \frac{\sin 
		\ub}{\sqrt{1 - \eb^2} (1 - \eb \cos \ub)} \ep \,.
\end{align}
Expanded to $\bigO(\xb^3)$ and $\bigO(\eb)$ this finally gives us
\begin{align}
	\phi =&\; \lab + 2 \eb \sin(\lb) + \xb (10-\nu) \eb \sin(\lb) \nonumber\\
		&+ \xb^2 \left( 52 - \frac{235\nu}{12} + \frac{\nu^2}{12} \right) \eb 
		\sin(\lb) \nonumber\\ 
		&- \xb^{5/2} \nu \left( \frac{128}{5} + \frac{888}{5} \eb \cos(\lb) 
		\right) \nonumber\\ 
		&+ \xb^3 \bigg( 292 + \left(-\frac{420131}{840} + \frac{287\pi^2}{32} 
		\right) \nu \nonumber\\
		&+ \frac{521\nu^2}{24} + \frac{\nu^3}{24} \bigg) \eb \sin(\lb) \,.
\end{align}
This is very similar to Eq.~(\ref{eq: phi solution}), but with the quantities
on the RHS replaced by their slowly evolving parts and with additional terms
at 2.5PN order.

\section{Hereditary Contributions}\label{sec: hereditary}

\subsection{Tail integrals}

Note that tail effects start appearing at 1.5PN order, and thus post-adiabatic
corrections to those will only enter the waveform at 4PN order and beyond. We
can thus neglect any radiation-reaction effects in this section and only
consider the conservative problem. At the end, we can then replace all
variables ($x$, $e$, $l$, $\lambda$) with their slowly evolving counterparts
($\xb$, $\eb$, $\lb$, $\lab$) to get the secular evolving amplitudes.

We now employ the quasi-Keplerian parametrization introduced in Sec.~\ref{sec: 
keplerian parametrization}. As we use the two angles $l$ and $\lambda$ to
parameterize the orbital motion, time derivatives of the source multipole
moments ($I_L$, $J_L$) can be calculated as
\begin{align}
	\frac{\ud}{\ud t} =&\; n \left( \frac{\ud}{\ud l}  + (1+k) 
		\frac{\ud}{\ud\lambda} \right) \,.
\end{align}

We use a low-eccentricity expansion to simplify expressions, so we expand
everything in powers of both $x$ and $e$. Inserting Eqs.~(\ref{eq: KE solution 
expanded}) into the source multipole moments (\ref{eq: I_L}--\ref{eq: J_L}),
and substituting those into the radiative moments (\ref{eq: U_L}--\ref{eq: 
V_L}) we can then easily calculate the spherical harmonic modes in terms of
$l$ and $\lambda$. We find, e.g., for the dominant $h^{22}_\tail$ mode
\begin{align}
	h^{22}_\tail =&\; \frac{8 G m \nu}{c^2 R} x^{5/2} \sqrt{\frac{\pi}{5}} 
		\frac{x^{3/2} c^3}{Gm} \nonumber\\
		&\times \int_{0}^{\infty} \ud\tau\, \ue^{-2\ui (\lambda - 
		\lambda(\tau))} \left[ \ln\left( \frac{\tau}{2\tau_0} \right) + 
		\frac{11}{12} \right] \nonumber\\
		&\times \bigg[ -8 + e \left( \frac{3}{2} \ue^{\ui (l - l(\tau))} - 
		\frac{81}{2} \ue^{-\ui (l - l(\tau))} \right) \bigg] \,,
\end{align}
where $l(\tau) = n\tau$ and $\lambda(\tau) = (1+k)n\tau$ and where we restrict 
ourselves to the leading post-Newtonian order and $\bigO(e)$. All other modes 
can be calculated similarly and be given as integrals over past history. These 
integrals can then be solved using the standard formulas
\begin{subequations}
\begin{align}
	\int_{0}^{\infty} \ud\tau\,\ue^{-\ui\omega \tau} =&\; -\frac{\ui}{\omega} 
		\,,\\
	\int_{0}^{\infty} \ud\tau\,\ue^{-\ui\omega \tau} \ln\left( 
		\frac{\tau}{2\tau_0} \right)=&\; \nonumber\\
		-\frac{1}{\omega} \bigg( \frac{\pi}{2} \sign{\omega} \;-&\; \ui 
		\left[\ln(2|\omega| \tau_0)	+ \eg \right] \bigg) \,,\\
	\int_{0}^{\infty} \ud\tau\,\ue^{-\ui\omega \tau} \ln^2\left( 
		\frac{\tau}{2\tau_0} \right)=&\; \nonumber\\
		-\frac{\ui}{\omega} \bigg( \frac{\pi^2}{6} +\bigg( \frac{\pi}{2} 
		\sign{\omega} \;-&\; \ui \left[ \ln(2|\omega| \tau_0) + \eg \right] 
		\bigg)^2 \bigg) \,.
\end{align}
\end{subequations}
Note that for terms of the form $\int \ud\tau\, \ue^{-\ui (\alpha \, l(\tau) + 
\beta \, \lambda(\tau))} [\dots]$ we have $\omega = n(\alpha + (1+k) \beta)$.

We are now able to give the tail contributions to the spherical harmonic modes
in terms of the parameters $x$, $e = e_t$ and the angles $\phi$ and $l$. The
modes have the following structure:
\begin{align}
	h^{\ell m}_\tail =&\; \frac{8 G m \nu}{c^2 R} x \sqrt{\frac{\pi}{5}} 
		\ue^{-\ui m\phi} H^{\ell m}_\tail \,.
\end{align}
The various contributions to, e.g., the $H^{22}_\tail$ mode are given to 
$\bigO(e)$ by
\begin{widetext}
\begin{subequations}\label{eq: h22-tail}
\begin{align}
	(H^{22}_\tail)_\textnormal{1.5PN} =&\; x^{3/2} \Bigg( 2 \pi + 6 \ui \ln 
		\left( \frac{x}{x_0'} \right) + e \bigg\{ \ue^{-\ui l} \left[ \frac{11 
		\pi}{4} + \frac{27 \ui}{2} \ln \left( \frac{3}{2} \right) + 
		\frac{33}{4} \ui \ln \left( \frac{x}{x_0'} \right) \right] \nonumber\\ 
		&+\ue^{\ui l} \left[\frac{13 \pi }{4} + \frac{3 \ui}{2} \ln (2) + 
		\frac{39}{4} \ui \ln \left( \frac{x}{x_0'} \right) \right] \bigg\} 
		\Bigg) \,,\\
	(H^{22}_\tail)_\textnormal{2.5PN} =&\; x^{5/2} \Bigg( \pi \left( 
		-\frac{107}{21} + \frac{34 \nu}{21} \right) + \left( -\frac{107 \ui}{7} 
		+ \frac{34 \ui \nu}{7} \right) \ln \left( \frac{x}{x_0'} \right) 
		\nonumber\\
		&+ e \bigg\{ \ue^{\ui l} \bigg[ -\frac{9 \ui}{2} + \pi \left( 
		\frac{229}{168} + \frac{61 \nu}{42} \right) + \left( \frac{473 \ui}{28} 
		- \frac{3 \ui \nu}{7} \right) \ln (2) + \left( \frac{229 \ui}{56} + 
		\frac{61 \ui \nu }{14} \right) \ln \left( \frac{x}{x_0'} \right) \bigg] 
		\nonumber\\
		&+ \ue^{-\ui l} \bigg[ -\frac{27 \ui}{2} + \pi \left( -\frac{1081}{168} 
		+ \frac{137 \nu}{42} \right) + \left( \frac{27 \ui}{4} + 9 \ui \nu 
		\right) \ln \left(\frac{3}{2} \right) \nonumber\\ 
		&+ \left( -\frac{1081 \ui}{56} + \frac{137 \ui \nu }{14} \right) \ln 
		\left( \frac{x}{x_0'} \right) \bigg] \bigg\} \Bigg) \,,\\
	(H^{22}_\tail)_\textnormal{3PN} =&\; x^3 \Bigg( -\frac{515063}{22050} + 
		\frac{428 \ui \pi }{105} + \frac{2 \pi^2}{3} + \left( -\frac{428}{35} + 
		12 \ui \pi \right) \ln \left( \frac{x}{x_0'} \right) - 18 \ln^2 \left( 
		\frac{x}{x_0'} \right) \nonumber\\
		&+ e \bigg\{ \ue^{-\ui l} \bigg[ -\frac{515063}{7200} + \frac{749 \ui 
		\pi}{60} + \frac{49 \pi^2}{24} + \left( -\frac{2889}{70} + \frac{81 \ui 
		\pi}{2} \right) \ln \left( \frac{3}{2} \right) - \frac{81}{2} \ln^2 
		\left( \frac{3}{2} \right) \nonumber\\
		&+ \left( -\frac{749}{20} + \frac{147 \ui \pi	}{4} - \frac{243}{2} 
		\ln \left( \frac{3}{2} \right) \right) \ln \left( \frac{x}{x_0'} 
		\right) - \frac{441}{8}\ln^2\left( \frac{x}{x_0'} \right) \bigg] 
		\nonumber\\
		&+ \ue^{\ui l} \bigg[ -\frac{14936827}{352800} + \frac{3103 \ui \pi 
		}{420} + \frac{29 \pi^2}{24} + \left( -\frac{107}{70} + \frac{3 \ui 
		\pi}{2} \right) \ln (2) + \frac{3}{2} \ln^2(2) \nonumber\\
		&+ \left( -\frac{3103}{140} + \frac{87 \ui \pi }{4} - \frac{9}{2} 
		\ln(2) \right) \ln \left( \frac{x}{x_0'} \right) -\frac{261}{8} \ln^2 
		\left( \frac{x}{x_0'} \right) \bigg] \bigg\} \Bigg) \,.
\end{align}
\end{subequations}
\end{widetext}
Here, $x_0'$ is related to the arbitrary constant $\tau_0$ by
\begin{align}
	x_0' &= \left( \frac{Gm}{c^3} \frac{\ue^{11/12 - \eg}}{4 \tau_0} 
		\right)^{2/3} \,.
\end{align}
We list expressions for all $h^{\ell m}_\tail$ modes in a supplemental
\emph{Mathematica} file.

\subsection{Memory integrals}

The nonlinear memory effect arises from the nonlogarithmic integrals in
Eqs.~(\ref{eq: U_L}); e.g., for the $\ell = 2$ modes we have
\begin{align}
	U_{ij}^\mem (t_r) =&\; -\frac{2G}{7c^5} \int_{-\infty}^{t_r} \ud\tau\; 
		I_{a\la i}^{(3)}(\tau) I_{j\ra a}^{(3)}(\tau) \,.
\end{align}
There are two types of memory arising from these integrals: DC (or ``direct 
current'') memory and oscillatory memory. The DC memory is a slowly increasing, 
nonoscillatory contribution to the gravitational-wave amplitude, entering at 
Newtonian order. This leads to a difference in the amplitude between early and 
late times:
\begin{align}
	\Delta h_\mem &= \lim_{t \rightarrow +\infty} h(t) - \lim_{t \rightarrow 
		-\infty} h(t) \,.
\end{align}
The oscillatory memory, on the other hand, is a normal periodic contribution
entering the gravitational-wave amplitude at higher PN order. In 
Refs.~\cite{arun-2004} and~\cite{blanchet-2008}, the authors give expressions 
for both leading-order DC and oscillatory memory in the circular limit. The 
calculation of DC memory has been extended to 3PN order for circular binaries
in Ref.~\cite{favata-2009} and to Newtonian order for eccentric binaries
in Ref.~\cite{favata-2011}. In this paper, we will only briefly discuss the 
leading-order contributions to the DC and oscillatory memory for eccentric 
binaries, such that we can compare our results to the circular limit
in Ref.~\cite{blanchet-2008}. The complete post-Newtonian corrections to the 
nonlinear memory are dealt with in a subsequent paper~\cite{ebersold-2019}, 
completing the hereditary contributions to the gravitational-wave amplitudes 
for nonspinning eccentric binaries.

Following the same steps as in the previous section, we can calculate the
derivatives of the source moments, and we find, e.g., for the $20$-mode:
\begin{align}
	h^{20}_\memdc =&\; \frac{256}{7} \frac{G m \nu}{c^2 R} \sqrt{ 
		\frac{\pi}{30}} \int_{-\infty}^{t_r} \ud t\, \left( 1 + \frac{313}{48} 
		e^2 \right) x^5 \,.
\end{align}
We find that all DC memory modes will consist of such integrals of the form
\begin{align}
	h^{\ell 0}_\memdc \propto&\; \int_{-\infty}^{t_r} \ud t\, x^p(t) \, e^q(t) 
		\,.
\end{align}
One can rewrite this as an integral over the eccentricity
\begin{align}\label{eq: hl0 mem integral}
	h^{\ell 0}_\memdc \propto&\; \int_{e_i}^{e(t_r)} \ud e\, 
        \left( \frac{\ud e}{\ud t} \right)^{-1} x^p(e) \, e^q \,,
\end{align}
where $e_i$ is some initial eccentricity at early times. Solving the evolution
equations~(\ref{eq: peters-mathews}) to leading order, we find
\begin{align}
	x(e) =&\; x_0 \left( \frac{e_0}{e} \right)^{12/19} \,,
\end{align}
where $x(e_0) = x_0$. We can insert this into Eq.~(\ref{eq: hl0 mem integral})
together with the evolution equation $\ud e/\ud t$ and integrate over $e$.
We then find DC memory at leading Newtonian order in the $20$-mode and
$40$-mode:
\begin{subequations}
\begin{align}
	h^{20}_\memdc =&\; \frac{8 G m \nu}{c^2 R} x \sqrt{\frac{\pi}{5}} \; 
		\frac{-5}{14 \sqrt{6}} \left\{ 1 - \left( \frac{e}{e_i} \right)^{12/19} 
		\right\} \,,\\
	h^{40}_\memdc =&\; \frac{8 G m \nu}{c^2 R} x \sqrt{\frac{\pi}{5}} \; 
		\frac{-1}{504 \sqrt{2}} \left\{ 1 - \left( \frac{e}{e_i} 
		\right)^{12/19} \right\} \,.
\end{align}
\end{subequations}

The time derivatives of the oscillatory modes are computed in the same way. We
find that they consist of integrals of the form
\begin{align}
	h^{\ell m}_\memosc \propto&\; \int_{-\infty}^{t_r} \ud t\, x^p(t) \, e^q(t) 
		\, \ue^{\ui (s \lambda + r l)} \,,
\end{align}
which can be integrated to give
\begin{align}
	h^{\ell m}_\memosc \propto&\; -\frac{\ui}{n ( r + (1+k) s)} x^p \, e^q \, 
		\ue^{\ui (s \lambda + r l)} \,.
\end{align}
Note that there are oscillatory memory contributions entering the waveform at
1.5, 2, 2.5 and 3PN order. We list here only the 2.5 and 3PN terms that have a
circular limit, as to compare our results to Ref.~\cite{blanchet-2008}. We 
refer to our follow-up work~\cite{ebersold-2019} for a complete treatment of 
nonlinear memory. The modes have the following structure:
\begin{align}
	h^{\ell m}_\memosc =&\; \frac{8 G m \nu}{c^2 R} x \sqrt{\frac{\pi}{5}} 
		\ue^{-\ui m\phi} H^{\ell m}_\memosc \,.
\end{align}
The various contributions to $\bigO(e)$ are:
\begin{subequations}
\begin{align}
	H^{31}_\memosc =&\; \frac{-121\, x^3 \nu \Delta}{45 \sqrt{14}} \left( 1 + e 
		\left\{ \frac{301}{242} \ue^{-\ui l} + \ue^{\ui l} \right\} \right) 
		\,,\\
	H^{33}_\memosc =&\; \frac{11\, x^3 \nu \Delta}{27 \sqrt{210}} \left( 1 + e 
		\left\{ \frac{9}{2} \ue^{-\ui l} + \frac{3}{22} \ue^{\ui l} \right\} 
		\right) \,,\\
	H^{44}_\memosc =&\; \frac{\ui\, x^{5/2} \nu}{9 \sqrt{35}} \left( 1 + e 
		\left\{ \frac{7}{5} \ue^{-\ui l} + 3 \ue^{\ui l} \right\} \right) \,,\\
	H^{51}_\memosc =&\; \frac{-13\, x^3 \nu \Delta}{63 \sqrt{385}} \left( 1 + e 
		\left\{ \frac{251}{208} \ue^{-\ui l} + \ue^{\ui l} \right\} \right) 
		\,,\\
	H^{53}_\memosc =&\; \frac{-x^3 \nu \Delta}{189 \sqrt{330}} \left( 1 + e 
		\left\{ \frac{201}{16} \ue^{\ui l} - \frac{369}{32} \ue^{-\ui l} 
		\right\} \right) \,,\\
	H^{55}_\memosc =&\; \frac{9\, x^3 \nu \Delta}{35 \sqrt{66}} \left( 1 + e 
		\left\{ \frac{2285}{1296} \ue^{-\ui l} + \frac{985}{288} \ue^{\ui l} 
		\right\} \right)\,.
\end{align}
\end{subequations}
%

%\vspace{-.01cm}
\section{Constructing the full 3PN-accurate waveform}\label{sec: full waveform}

We now want to construct the full 3PN-accurate waveform valid during the 
inspiral of a binary system. We begin by adding up the two contributions to the 
spherical harmonic modes:
\begin{align}
	h^{\ell m} &= (h^{\ell m})_\inst + (h^{\ell m})_\hered \,.
\end{align}
Note that we are still missing some memory contributions. These will be 
computed in full in our follow-up work~\cite{ebersold-2019}, and we will give 
expressions for the full waveform including memory there.

\subsection{Instantaneous parts}

The instantaneous parts $(h^{\ell m})_\inst$ of the spherical harmonic modes
for compact binaries in elliptical orbits have already been calculated to the
third post-Newtonian order in Ref.~\cite{mishra-2015}, although the results do 
not include post-adiabatic corrections to the quasi-Keplerian parametrization.
They are given in terms of the constants of motion $x$ and $e = e_t$ and
parametrized by the eccentric anomaly $u$. We will rewrite these in terms of
the mean anomaly $l$ by using the solution to the Kepler equation~(\ref{eq: u 
solution}). This gives us expressions for the instantaneous contributions to 
the different modes in terms of the post-Newtonian parameter $x$ and the time 
eccentricity $e$, parametrized by the angles $\phi$ and $l$. The modes again 
have the following structure:
\begin{align}\label{eq: hlm inst}
	h^{\ell m}_\inst =&\; \frac{8 G m \nu}{c^2 R} x \sqrt{\frac{\pi}{5}} 
		\ue^{-\ui m\phi} H^{\ell m}_\inst \,.
\end{align}
The various contributions to, e.g., the $H^{22}_\inst$ mode are given to
$\bigO(e)$ by
\begin{widetext}
\begin{subequations}\label{eq: h22 inst}
\begin{align}
	(H^{22}_\inst)_\textnormal{Newt} =&\; 1 + e \bigg\{ \frac{1}{4} 
		\ue^{-\ui l} + \frac{5}{4} \ue^{\ui l} \bigg\} \,,\\
	(H^{22}_\inst)_\textnormal{1PN} =&\; x \Bigg( -\frac{107}{42} + \frac{55 
		\nu}{42} + e \bigg\{ \ue^{-\ui l} \left[ -\frac{257}{168} + 
		\frac{169 \nu}{168} \right] + \ue^{\ui l} \left[ -\frac{31}{24} + 
		\frac{35 \nu}{24} \right] \bigg\} \Bigg) \,,\\
	(H^{22}_\inst)_\textnormal{2PN} =&\; x^2 \Bigg( -\frac{2173}{1512} - 
		\frac{1069 \nu}{216} + \frac{2047 \nu^2}{1512} + e \bigg\{ 
		\ue^{\ui l} \left[ -\frac{2155}{252} - \frac{1655 \nu}{672} + 
		\frac{371 \nu^2}{288} \right] \nonumber\\
		&+ \ue^{-\ui l} \left[ -\frac{4271}{756} - \frac{35131 \nu}{6048} + 
		\frac{421 \nu^2}{864} \right] \bigg\} \Bigg) \,,\\
	(H^{22}_\inst)_\textnormal{2.5PN} =&\; -x^{5/2} \ui \nu \Bigg( \frac{56}{5} 
		+ e \bigg\{ \frac{7817}{420} \ue^{\ui l} + \frac{2579}{84} 
		\ue^{-\ui l} \bigg\} \Bigg) \,,\\
	(H^{22}_\inst)_\textnormal{3PN} =&\; x^3 \Bigg( \frac{761273}{13200} + 
		\left( -\frac{278185}{33264} + \frac{41 \pi^2}{96} \right) \nu - 
		\frac{20261 \nu^2}{2772} + \frac{114635 \nu^3}{99792} + \frac{856}{105} 
		\ln \left( \frac{x}{x_0} \right) \nonumber\\
		&+ e \bigg\{ \ue^{\ui l} \left[ \frac{6148781}{75600} + \left( 
		-\frac{199855}{3024} + \frac{41 \pi^2}{48} \right) \nu - \frac{9967 
		\nu^2}{1008} + \frac{35579 \nu^3}{36288} + \frac{3103}{210} \ln \left( 
		\frac{x}{x_0} \right) \right] \nonumber\\
		&+ \ue^{-\ui l} \left[ \frac{150345571}{831600} + \left( 
		-\frac{121717}{20790} - \frac{41 \pi^2}{192} \right) \nu - \frac{86531 
		\nu^2}{8316} - \frac{33331 \nu^3}{399168} + \frac{749}{30} \ln \left( 
		\frac{x}{x_0} \right) \right] \bigg\} \Bigg) \,,
\end{align}
\end{subequations}
\end{widetext}
where $x_0 = Gm/(c^3 \tau_0)$ is related to $x_0'$ by
\begin{align}\label{eq: logx0 relation}
	\ln x_0' &= \frac{11}{18} -\frac{2}{3}\eg - \frac{4}{3} \ln 2 + \frac{2}{3} 
		\ln x_0 \,.
\end{align}

\subsection{Post-adiabatic corrections}

We now move to include post-adiabatic corrections into the waveform. As already 
mentioned in Sec.~\ref{sec: hereditary}, post-adiabatic corrections to the 
hereditary contributions will only enter at 4PN. We are thus left with 
computing the corrections to the instantaneous contributions as described in 
Sec.~\ref{sec: phasing}. Schematically, the substitutions in Eq.~(\ref{eq: 
phasing subst}) may be described as
\begin{align}
	h^{\ell m}(&x, e, l, \lambda) \nonumber\\
	&\Downarrow \nonumber\\
	h^{\ell m}(\xb + \xp, \eb &+ \ep, \lb + \lp, \lab + \lap) \nonumber\\
	&\Downarrow \nonumber\\
	h^{\ell m}(\xb, \eb, \lb, \lab) + \bigg\{ \frac{\partial h^{\ell 
		m}}{\partial x} \xp \,+\,& \frac{\partial h^{\ell m}}{\partial e} \ep + 
		\frac{\partial h^{\ell m}}{\partial l} \lp + \frac{\partial h^{\ell 
		m}}{\partial \lambda} \lap \bigg\} \nonumber\\
	&\Downarrow \nonumber\\
	h^{\ell m}(\xb, \eb , \lb, \lab) \,+&\, \frac{1}{c^5} \, h^{\ell m}_\post 
		(\xb, \eb, \lb, \lab) \,.
\end{align}
In particular, we only need to make these substitutions at leading Newtonian 
and 0.5PN order. At higher orders, we simply replace the variables ($x$, $e$, 
$l$, $\lambda$) by their secular evolving parts ($\xb$, $\eb$, $\lb$, $\lab$) 
to get the secular evolving waveform. 

The post-adiabatic contributions to the different modes in terms of the secular
evolving parameters $\xb$ and $\eb$, parametrized by the angles $\phi$ and
$\lb$, have the following form:
\begin{align}
	h^{\ell m}_\post =&\; \frac{8 G m \nu}{c^2 R} \xb \sqrt{\frac{\pi}{5}} 
		\ue^{-\ui m\phi} H^{\ell m}_\post \,.
\end{align}
For example, the $H^{22}_\post$ mode, that arises from including the 
post-adiabatic corrections in $(H^{22}_\inst)_\textnormal{Newt}$, is given by
\begin{align}\label{eq: h22-post-ad}
	H^{22}_\post =&\; \nonumber\\
		\frac{192}{5} & \xb^{5/2} \ui \nu \Bigg( 1 + \eb \bigg\{ \frac{401}{72} 
		\ue^{-\ui\lb} + \frac{293}{72} \ue^{\ui\lb} \bigg\} \Bigg) \,.
\end{align}

We can combine these post-adiabatic contributions with the instantaneous ones 
to get the full secular evolving instantaneous waveform in terms of the 
variables ($\xb$, $\eb$, $\lb$, $\lab$). The result has again the following 
form:
\begin{align}
	h^{\ell m}_\inst =&\; \frac{8 G m \nu}{c^2 R} \xb \sqrt{\frac{\pi}{5}} 
		\ue^{-\ui m\phi} H^{\ell m}_\inst \,.
\end{align}
In e.g.~the $H^{22}_\inst$ mode we find that the only term that is modified is 
the 2.5PN order:
\begin{align}
	&(H^{22}_\inst)_\textnormal{2.5PN} =\nonumber\\ 
		&\quad\quad 
		-\xb^{5/2} \ui \nu \Bigg( 24 + \eb \bigg\{ \frac{43657}{420} 
		\ue^{\ui\lb} + \frac{1013}{140} \ue^{-\ui\lb} \bigg\} \Bigg) \,.
\end{align}
All other orders are exactly as in Eqs.~(\ref{eq: h22 inst}), but with ($x$, 
$e$, $l$, $\lambda$) replaced by ($\xb$, $\eb$, $\lb$, $\lab$) .

\subsection{Log cancellation}\label{sec: log cancel}

We observe that both instantaneous and tail terms still have some dependence on 
the arbitrary constant $x_0'$ (or $x_0$). We find that this dependence on 
$x_0'$ can be reabsorbed in a shift of the coordinate time $t$ 
\cite{blanchet-1996, arun-2004} through a redefinition of the mean anomaly as
\begin{align}
	\xi &= \lb - \frac{3GM}{c^3} \bar{n} \ln \Big( \frac{\xb}{x_0'} \Big) \,,
\end{align}
where $M = m (1 - \nu \xb / 2)$ is the ADM mass. Note that there are no 
post-adiabatic corrections to $n$ and $x$ here, as phasing effects would only 
enter at $1.5+2.5$PN order. This also means that both $\xi$ and $\lb$ follow 
the same evolution, i.e., $\ud\xi/\ud t = \ud\lb/\ud t = \bar{n}$, and they 
only differ by a constant factor. To simplify the final expressions, we also 
introduce a redefined phase $\psi$ such that Eq.~(\ref{eq: quasi-keplerian phi 
bar}) gives the relation between $\xi$ and $\psi$:
\begin{align}
	\psi =\;& \lab_\xi + \bar{W}_\xi + \lap_\xi + (\vp_\xi - \lp_\xi) \,.
\end{align}
Here
\begin{align}
	\lab_\xi =&\; \lab - \frac{3GM}{c^3} (1 + \bar{k}) \bar{n} \ln \Big( 
		\frac{\xb}{x_0'} \Big) \,, 
\end{align}
is the phase $\lab$ evaluated at the shifted time defined by $\xi$, and
$\bar{W}_\xi$, $\lap_\xi$, $\vp_\xi$, and $\lp_\xi$ are defined as in
Eq.~(\ref{eq: quasi-keplerian phi bar}), but with $\lb$ replaced by $\xi$.
From this, we can easily deduce that
\begin{align}
	\psi =\;& \phi + \sum_{s=1}^{\infty} \frac{1}{s!} \Bigg[ \left( \xi - \lb 
		\right)^s \left( \frac{\ud}{\ud\lb} \right)^s \nonumber\\
		&+ \left( \lab_\xi - \lab \right)^s \left( \frac{\ud}{\ud\lab} 
		\right)^s \Bigg] \phi \,. 
\end{align}
Note that the phase $\psi$ does not have the same geometric interpretation as
$\phi$. Expanding these equations to $\bigO(\xb^3)$ and $\bigO(\eb)$, we find
\begin{subequations}
\begin{align}
	\lb =&\; \xi + 3 \left( \xb^{3/2} - \xb^{5/2} \left( 3 + \frac{\nu}{2} 
		\right) \right) \ln \Big( \frac{\xb}{x_0'} \Big) \,,\\
	\phi =&\; \psi + \bigg( \xb^{3/2} \left( 3 + 6 \eb \cos(\xi) \right) 
		\nonumber\\
		&+ \xb^{5/2} \left( -\frac{3\nu}{2} + 6 \eb (2 - \nu) \cos(\xi) \right) 
		\bigg) \ln \Big( \frac{\xb}{x_0'} \Big) \nonumber\\
		&- 9 \xb^3 \eb \sin(\xi) \ln^2 \Big( \frac{\xb}{x_0'} \Big) \,.
\end{align}
\end{subequations}

This redefinition of the time coordinate results in the cancellation of all log
terms involving the arbitrary constant $x_0'$.

\subsection{Full waveform}

The full waveform in terms of the redefined angles $\xi$ and $\psi$ -- minus 
some memory contributions -- has the following form:
\begin{align}
	h^{\ell m} =&\; \frac{8 G m \nu}{c^2 R} \xb \sqrt{\frac{\pi}{5}} 
		\ue^{-\ui m\psi} H^{\ell m} \,.
\end{align}
The various contributions to, e.g., the $H^{22}$ mode are given to $\bigO(\eb)$ 
by
\begin{widetext}
\begin{subequations}\label{eq: Hlm inst+hered}
\begin{align}
	H^{22}_\textnormal{Newt} =&\; 1 + \eb \bigg\{ \frac{1}{4} \ue^{-\ui\xi} + 
		\frac{5}{4} \ue^{\ui\xi} \bigg\} \,,\\
	H^{22}_\textnormal{1PN} =&\; \xb \Bigg( -\frac{107}{42} + \frac{55 \nu}{42} 
		+ \eb \bigg\{ \ue^{-\ui\xi} \left[ -\frac{257}{168} + \frac{169 
		\nu}{168} \right] + \ue^{\ui\xi} \left[ -\frac{31}{24} + \frac{35 
		\nu}{24} \right] \bigg\} \Bigg) \,,\\
	H^{22}_\textnormal{1.5PN} =&\; \xb^{3/2} \Bigg( 2 \pi + \eb	\bigg\{ 
		\ue^{-\ui \xi} \left[ \frac{11 \pi }{4} + \frac{27 \ui}{2} \ln \left( 
		\frac{3}{2} \right) \right] + \ue^{\ui \xi} \left[\frac{13 \pi }{4} + 
		\frac{3 \ui}{2} \ln(2) \right] \bigg\} \Bigg) \,,\\
	H^{22}_\textnormal{2PN} =&\; \xb^2 \Bigg( -\frac{2173}{1512} - \frac{1069 
		\nu}{216} + \frac{2047 \nu^2}{1512} + \eb \bigg\{ \ue^{\ui\xi} \left[ 
		-\frac{2155}{252} - \frac{1655 \nu}{672} + \frac{371 \nu^2}{288} 
		\right] \nonumber\\
		&+ \ue^{-\ui\xi} \left[ -\frac{4271}{756} - \frac{35131 \nu}{6048} + 
		\frac{421 \nu^2}{864} \right] \bigg\} \Bigg) \,,\\
	H^{22}_\textnormal{2.5PN} =&\; \xb^{5/2} \Bigg( -\frac{107 \pi}{21} + 
		\left( -24 \ui + \frac{34 \pi}{21} \right) \nu \nonumber\\
		&+ \eb \bigg\{ \ue^{\ui \xi} \bigg[ -\frac{9 \ui}{2} + \frac{229 
		\pi}{168} + \left( -\frac{43657 \ui}{420} + \frac{61 \pi}{42} \right) 
		\nu + \left( \frac{473 \ui}{28} - \frac{3 \ui \nu }{7} \right) \ln (2) 
		\bigg] \nonumber\\
		&+ \ue^{-\ui \xi} \bigg[ -\frac{27 \ui}{2} -\frac{1081 \pi}{168} + 
		\left( -\frac{1013 \ui}{140} + \frac{137 \pi}{42} \right) \nu + \left( 
		\frac{27 \ui}{4} + 9 \ui \nu \right) \ln \left( \frac{3}{2} \right) 
		\bigg] \bigg\} \Bigg) \,,\\
	H^{22}_\textnormal{3PN} =&\; \xb^3 \Bigg( \frac{27027409}{646800} + 
		\frac{428 \ui \pi}{105} + \frac{2 \pi^2}{3} - \frac{856 \eg}{105} + 
		\left( -\frac{278185}{33264} + \frac{41 \pi^2}{96} \right) \nu - 
		\frac{20261 \nu^2}{2772} + \frac{114635 \nu^3}{99792} \nonumber\\
		&- \frac{1712 \ln(2)}{105} - \frac{428 \ln(\xb)}{105} \nonumber\\
		&+ \eb \bigg\{ \ue^{-\ui \xi} \bigg[ \frac{219775769}{1663200} + 
		\frac{749 \ui \pi}{60} + \frac{49 \pi^2}{24} - \frac{749 \eg}{30} + 
		\left( -\frac{121717}{20790} - \frac{41 \pi^2}{192}\right) \nu - 
		\frac{86531 \nu^2}{8316} - \frac{33331 \nu^3}{399168} \nonumber\\
		&+ \left( -\frac{2889}{70} + \frac{81 \ui \pi}{2}\right) \ln \left( 
		\frac{3}{2} \right) - \frac{81}{2} \ln^2 \left( \frac{3}{2} \right) - 
		\frac{749 \ln(2)}{15} - \frac{749 \ln(\xb)}{60} \bigg] \nonumber\\
		&+ \ue^{\ui \xi} \bigg[ \frac{55608313}{1058400} + \frac{3103 \ui 
		\pi}{420} + \frac{29 \pi^2}{24} - \frac{3103 \eg}{210} + \left( 
		-\frac{199855}{3024} + \frac{41 \pi^2}{48} \right) \nu -\frac{9967 
		\nu^2}{1008} + \frac{35579 \nu^3}{36288} \nonumber\\
		&+ \left( -\frac{6527}{210} + \frac{3 \ui \pi}{2}\right) \ln(2) + 
		\frac{3 \ln^2(2)}{2} - \frac{3103 \ln(\xb)}{420} \bigg] \bigg\} \Bigg) 
		\,.
\end{align}
\end{subequations}
\end{widetext}
For completeness all equations relating the different angles $\lb$, $\lab$, 
$\xi$ and $\psi$ are listed in Appendix~\ref{sec: quasi-kepl relations}.

\subsection{Quasi-Circular limit}\label{sec: circular}

We now check our results against those in Ref.~\cite{blanchet-2008} in the
quasi-circular limit. Note that the eccentricity is not a gauge-independent
quantity and one thus has to be careful when talking about the circular limit.
For a thorough discussion on different eccentricity parameters and
discrepancies between them we refer to Refs.~\cite{loutrel-2018, loutrel-2019}.

Normally, one uses the orbital averaged description for the evolution of $x$
and $e$, where one finds that the evolution equations~(\ref{eq: peters-mathews})
drive the eccentricity to zero during the inspiral. When introducing
post-adiabatic corrections, this will not be true anymore, as the eccentricity
is split into a orbital averaged part $\eb$ and a periodic oscillatory part
$\ep$. The orbital averaged part $\eb$ will still follow the same evolution
equations~(\ref{eq: peters-mathews}) and thus be driven to zero, but the 
periodic variations $\ep$ will generally grow larger as the binary inspirals. As
discussed in Ref.~\cite{loutrel-2019}, the orbital averaged description also 
breaks down in the late inspiral, failing to capture a secular growth in the
eccentricity observed when directly integrating the two-body equations of
motion.

In our case, it is reasonable to consider the circular limit as the limit where
$\xb \rightarrow x$ and $\eb \rightarrow 0$, with $x$ being the standard
circular frequency parameter. Then, the evolution equations~(\ref{eq: 
peters-mathews}) reduce to the usual circular evolution equation
\begin{align}
	\dot{x} &= \frac{64c^3 \nu}{5Gm} x^5 + \bigO(x^6)\,.
\end{align}
In this limit our redefined phase $\psi$ reduces to
\begin{align}
	\psi|_{\eb=0} &= \phi - 3 \left(1 - \frac{\nu x}{2}\right) x^{3/2} \ln 
		\Big( \frac{x}{x_0'} \Big) \,,
\end{align}
which matches exactly the phase $\psi$ used in Ref.~\cite{blanchet-2008}. We can
thus directly compare our results to the circular limit by setting $\eb = 0$
and $\xb|_{\eb=0} = x$. We find, e.g., for the $h^{22}$ mode
\begin{align}
	h^{22} = \frac{8Gm\nu}{c^2 R} x \sqrt{\frac{\pi}{5}} \ue^{-2\ui\psi} H^{22} 
	\,,
\end{align}
\begin{widetext}
\begin{align}
	H^{22} =\;& 1 + x \left( -\frac{107}{42} + \frac{55\nu}{42} \right) + 2\pi 
	x^{3/2} + x^2 \left( -\frac{2173}{1512} - \frac{1069\nu}{216} + 
		\frac{2047\nu^2}{1512} \right) + x^{5/2} \left( -\frac{107\pi}{21} + 
		\left( -24\ui + \frac{34\pi}{21} \right) \nu \right) \nonumber\\
		&+ x^3 \bigg( \frac{27027409}{646800} + \frac{428\ui\pi}{105} + \frac{2 
		\pi^2}{3} - \frac{856 \eg}{105} + \left( -\frac{278185}{33264} + 
		\frac{41\pi^2}{96} \right) \nu - \frac{20261 \nu^2}{2772} + 
		\frac{114635\nu^2}{99792} \nonumber\\
		&- \frac{1712}{105} \ln(2) - \frac{428}{105} \ln(x) \bigg) \,.
\end{align}
\end{widetext}
This matches Eq.~(9.4a) of Ref.~\cite{blanchet-2008}. Similarly, we can compare 
the other modes and find perfect agreement in all of them.

\section{Conclusion}\label{sec: summary}

In this work, we computed the tail contributions to the 3PN-accurate
gravitational waveform from nonspinning compact binaries on eccentric orbits.
This extends the work on instantaneous contributions in Ref.~\cite{mishra-2015} 
and will be completed with the memory contributions in a follow-up
paper~\cite{ebersold-2019}. We also include post-adiabatic corrections to the
quasi-Keplerian parametrization when combining our tail results with the
instantaneous ones, giving us the full waveform (neglecting memory) that can
be compared to the circular one in the limit $e \rightarrow 0$. The tail
contributions to the $h^{22}$ mode are given at 3PN order and to $\bigO(e)$ in
Eq.~(\ref{eq: h22-tail}), the post-adiabatic corrections in Eq.~(\ref{eq: 
h22-post-ad}). All other $h^{\ell m}$ modes up to $\ell = 5$ are listed in
the supplemental \emph{Mathematica} notebook~\cite{supplement}. To reiterate, 
all results are in MH coordinates, which differ from the SH coordinates at 3PN 
order.

Note that the instantaneous results in Ref.~\cite{mishra-2015} can be applied to
binary systems of arbitrary eccentricities, while the tail results presented
here are calculated in a small eccentricity expansion. This is due to the
complicated tail integrals over past history, which can only be analytically
calculated when decomposing the integrand into harmonics of the orbital
timescale using an eccentricity expansion. This means that our results are not
applicable for large eccentricities $e \sim 1$, though they might give
accurate results for moderate eccentricities $e \sim 0.4$ when combined with
orbital evolution equations that are not expanded in eccentricity; see, e.g., 
Ref.~\cite{klein-2018}.

\acknowledgments

We thank Riccardo Sturani for a first review. Y.~B. is supported by the Swiss 
National Science Foundation and a Forschungskredit of the University of Zurich, 
Grant No.~FK-18-084. Y.~B. would like to acknowledge the hospitality of the 
Institut d’Astrophysique de Paris during the final stages of this collaboration.

\appendix

\begin{widetext}

\section{Integral relations}\label{sec: integrals}

We calculate the average over one period as
\begin{align}
	\la F \ra &= \frac{1}{T} \int_{0}^{T} F(t) \ud t = \frac{1}{2\pi} 
		\int_{0}^{2\pi} F(l) \ud l = \frac{1}{2\pi} \int_{0}^{2\pi} F(u) \chi
        \ud u \,,
\end{align}
where $\chi = 1 - e \cos u$. Some helpful integration formulas are
\begin{subequations}
\begin{align}
	\frac{1}{2\pi} \int_{0}^{2\pi} \frac{\ud u}{\chi^{N+1}} &= \frac{1}{(1 - 
		e^2)^{(N+1)/2}} P_N \left( \frac{1}{\sqrt{1 - e^2}} \right) \,,\\
	\frac{1}{2\pi} \int_{0}^{2\pi} \chi \ln (\chi) \ud u &= 1 - \sqrt{1 - e^2} 
		+ \ln \bigg( \frac{1 + \sqrt{1 - e^2}}{2} \bigg) \,.
\end{align}
\end{subequations}

The zero-average primitive of a function $F(l)$ can be calculated as
\begin{align}
	\int F &= \int_{0}^{l} F(l') \ud l' - \frac{1}{2\pi} \int_{0}^{2\pi} \ud l 
		\int_{0}^{l} F(l') \ud l' = \int_{0}^{l} F(l') \ud l' - \frac{1}{2\pi} 
		\int_{0}^{2\pi} (2\pi - l') F(l') \ud l' \,.
\end{align}

\section{Quasi-Keplerian relations}\label{sec: quasi-kepl relations}

We list here again all equations relating the different angles $\lb$, $\lab$,
$\xi$, and $\psi$ to the actual time coordinate $t$:
\begin{subequations}
\begin{align}
	\xi =&\; \lb - \frac{3GM}{c^3} \bar{n} \ln \Big( \frac{\xb}{x_0'} \Big) 
		\,,\\
	\lab_\xi =&\; \lab - \frac{3GM}{c^3} (1 + \bar{k}) \bar{n} \ln \Big( 
		\frac{\xb}{x_0'} \Big) 
		= \lab - 3 \left(1 - \frac{\nu \xb}{2}\right) \xb^{3/2} \ln \Big( 
		\frac{\xb}{x_0'} \Big) \,,\\
	\psi =&\; \lab_\xi + \bar{W}_\xi + \lap_\xi + (\vp_\xi - \lp_\xi) 
		\nonumber\\
		=&\; \lab_\xi + 2 \eb \sin(\xi) + \frac{5}{4} \eb^2 \sin(2 \xi)+\xb 
		\left( (10- \nu) \eb \sin(\xi) + \left( \frac{31}{4} - \nu \right) 
		\eb^2 \sin(2 \xi) \right) \nonumber\\
		&+ \xb^2 \left( \frac{1}{12} \left( 624 - 235 \nu + \nu^2 \right) \eb 
		\sin(\xi) + \frac{1}{24} \left( 969 - 326 \nu + 2 \nu^2 \right) \eb^2 
		\sin(2 \xi) \right) \nonumber\\
		&- \xb^{5/2} \nu \left( \frac{128}{5} + \frac{888}{5} \eb \cos (\xi) + 
		\frac{1}{45} \eb^2 (10728 + 8935 \cos (2 \xi)) \right) \nonumber\\
		&+ \xb^3 \bigg( \left( 292 + \left( -\frac{420131}{840} + \frac{287 
		\pi^2}{32} \right) \nu + \frac{521 \nu^2}{24} + \frac{\nu^3}{24} 
		\right) \eb \sin(\xi) \nonumber\\
		&+ \frac{1}{168} \left( 35868 + \left( -55548+861 \pi^2 \right) \nu + 
		1925 \nu^2 + 28 \nu^3 \right) \eb^2 \sin(2 \xi) \bigg) \,,
\end{align}
\end{subequations}
where
\begin{subequations}
\begin{align}
	\xi =&\; \ub_\xi - \eb \sin \ub_\xi + \left( \bar{g}_{4t} + \bar{g}_{6t} 
		\right)(\vb_\xi - \ub_\xi) + \left( \bar{f}_{4t} + \bar{f}_{6t} 
		\right)\sin \vb_\xi + \bar{i}_{6t} \sin (2\vb_\xi) + \bar{h}_{6t} 
		\sin(3\vb_\xi) \,,\\
	\ub_\xi =&\; \xi + \eb \sin(\xi) + \frac{1}{2} \eb^2 \sin(2 \xi) + \xb^2 
		\left( \frac{1}{8} \left( -60 + 9 \nu + \nu^2 \right) \eb \sin(\xi) + 
		\frac{3}{16} ( -5 + \nu) (10 + \nu) \eb^2 \sin(2 \xi) \right) 
		\nonumber\\
		&+ \xb^3 \bigg( \left( -55 + \frac{104593 \nu}{1680} + \frac{3 
		\nu^2}{4} + \frac{\nu^3}{24} \right) \eb \sin(\xi) \nonumber\\
		&+ \left( -\frac{315}{4} + \left( \frac{229219}{3360} + \frac{41 
		\pi^2}{256} \right) \nu + \frac{53 \nu^2}{8} - \frac{3 \nu^3}{16} 
		\right) \eb^2 \sin(2 \xi) \bigg) \,,\\
	\vb_\xi =&\; 2 \arctan \left[\left( \frac{ 1 + \eb_{\phi} }{1 - \eb_{\phi}} 
		\right)^{1/2} \tan \frac{\ub_\xi}{2} \right] \nonumber\\
		=&\; \xi + 2 \eb \sin(\xi) + \frac{5}{4} \eb^2 \sin(2 \xi) + \xb \left( 
		(4 - \nu) \eb \sin(\xi) + (4- \nu) \eb^2 \sin(2 \xi) \right) \nonumber\\
		&+ \xb^2 \left( \frac{1}{12} \left( 156 - 31 \nu + \nu^2 \right) \eb 
		\sin(\xi) + \frac{1}{24} \left( 273 - 101 \nu + 11 \nu^2 \right) \eb^2 
		\sin(2 \xi) \right) \nonumber\\
		&+ \xb^3 \bigg( \left( 64 + \left( -\frac{106181}{840} + \frac{41 
		\pi^2}{32} \right) \nu + \frac{11 \nu^2}{24} + \frac{\nu^3}{24} \right) 
		\eb \sin(\xi) \nonumber\\
		&+ \left( \frac{155}{4} + \left( -\frac{169649}{1680} + \frac{369 
		\pi^2}{256} \right) \nu + \frac{49 \nu^2}{6} - \frac{5 \nu^3}{24} 
		\right) \eb^2 \sin(2 \xi) \bigg) \,,\\
	\bar{W}_\xi =&\; (1 + \bar{k})(\vb_\xi - \xi) + \left( \bar{f}_{4\phi} + 
		\bar{f}_{6\phi} \right) \sin(2\vb_\xi) + \left( \bar{g}_{4\phi} + 
		\bar{g}_{6\phi} \right) \sin(3\vb_\xi) + \bar{i}_{6\phi}\sin(4\vb_\xi) 
		+ \bar{h}_{6\phi} \sin(5\vb_\xi) \nonumber\\
		=&\; 2 \eb \sin(\xi) + \frac{5}{4} \eb^2 \sin(2 \xi)+\xb \left( (10 - 
		\nu) \eb \sin(\xi) + \left( \frac{31}{4} - \nu \right) \eb^2 \sin(2 
		\xi) \right) \nonumber\\
		&+ \xb^2 \left( \frac{1}{12} \left( 624 - 235 \nu + \nu^2 \right) \eb 
		\sin(\xi) + \frac{1}{24} \left( 969 - 326 \nu + 2 \nu^2 \right) \eb^2 
		\sin(2 \xi) \right) \nonumber\\
		&+ \xb^3 \bigg( \left( 292 + \left( -\frac{420131}{840} + \frac{287 
		\pi^2}{32} \right) \nu + \frac{521 \nu^2}{24} + \frac{\nu^3}{24} 
		\right) \eb \sin(\xi) \nonumber\\
		&+ \frac{1}{168} \left( 35868 + \left( -55548+861 \pi^2 \right) \nu + 
		1925 \nu^2 + 28 \nu^3 \right) \eb^2 \sin(2 \xi) \bigg) \,,\\
	\lap_\xi =&\; -\nu \xb^{5/2} \bigg[ \frac{296}{3} \eb \cos(\xi) + 
		\frac{199}{5} \eb^2 \cos(2\xi) \bigg] + \bigO(\xb^{7/2}) \,,\\
	\lp_\xi =&\; -\nu \xb^{5/2} \bigg[ \frac{64}{5\eb} \cos(\xi) + 
		\frac{352}{15} \cos(2\xi) + \eb \left( \frac{1654}{15} \cos(\xi) + 
		\frac{358}{9} \cos(3\xi) \right) \nonumber\\
		&+ \eb^2 \left( \frac{694}{15} \cos(2\xi) + \frac{1289}{20} \cos(4\xi) 
		\right) \bigg] + \bigO(\xb^{7/2}) \,, \\
	\up_\xi =&\; -\nu \xb^{5/2} \bigg[ \frac{64}{5} + \frac{64}{5\eb} \cos(\xi) 
		+ \frac{352}{15} \cos(2\xi) + \eb \left( \frac{2198}{15} \cos(\xi) + 
		\frac{358}{9} \cos(3\xi) \right) \nonumber\\
		&+ \eb^2 \left( \frac{340}{3} + \frac{5366}{45} \cos(2\xi) + 
		\frac{1289}{20} \cos(4\xi) \right) \bigg] + \bigO(\xb^{7/2}) \,,\\
	\ep_\xi =&\; -\nu \xb^{5/2} \bigg[ \frac{64}{5} \sin(\xi) + \frac{352}{15} 
		\eb \sin(2\xi) + \eb^2 \left( \frac{1138}{15} \sin(\xi) + \frac{358}{9} 
		\sin(3\xi) \right) \bigg] + \bigO(\xb^{7/2}) \,,\\
	\vp_\xi =&\; \frac{\sqrt{1 - \eb^2}}{1 - \eb \cos \ub_\xi} \up_\xi + 
		\frac{\sin \ub_\xi}{\sqrt{1 - \eb^2} (1 - \eb \cos \ub_\xi)} \ep 
		\nonumber\\
		=&\; -\nu \xb^{5/2} \bigg[ \frac{128}{5} + \frac{64}{5\eb} \cos(\xi) + 
		\frac{352}{15} \cos(2\xi) + \eb \left( \frac{946}{5} \cos(\xi) + 
		\frac{358}{9} \cos(3\xi) \right) \nonumber\\
		&+ \eb^2 \left( \frac{1192}{5} + \frac{9226}{45} \cos(2\xi) + 
		\frac{1289}{20} \cos(4\xi) \right) \bigg] + \bigO(\xb^{7/2}) \,,
\end{align}
\end{subequations}
where $\lb$ and $\lab$ are given by Eqs.~(\ref{eq: secular l and la}) and where
$\bar{f}_{4\phi}$, $\bar{f}_{6\phi}$, $\bar{g}_{4\phi}$, $\bar{g}_{6\phi}$,
$\bar{i}_{6\phi}$, $\bar{h}_{6\phi}$, $\bar{g}_{4t}$, $\bar{g}_{6t}$,
$\bar{f}_{4t}$, $\bar{f}_{6t}$, $\bar{i}_{6t}$, and $\bar{h}_{6t}$ are the
slowly evolving orbital functions given in Ref.~\cite{memmesheimer-2004}, with 
$x$ and $e$ being replaced by $\xb$ and $\eb$.

\end{widetext}

\bibliographystyle{apsrev4-1}
\bibliography{tail-paper}

\end{document}